\documentclass[12pt]{article}
\usepackage{amsmath,amssymb}
\usepackage{graphicx,psfrag,epsf}
\usepackage{enumerate}
\usepackage{natbib}
\usepackage{url} 
\usepackage{mathtools}
\usepackage{comment}
\usepackage{soul}
\usepackage{amsbsy}

\usepackage{scalerel,stackengine}
\stackMath
\newcommand\reallywidehat[1]{%
\savestack{\tmpbox}{\stretchto{%
  \scaleto{%
    \scalerel*[\widthof{\ensuremath{#1}}]{\kern-.6pt\bigwedge\kern-.6pt}%
    {\rule[-\textheight/2]{1ex}{\textheight}}
  }{\textheight}%
}{0.5ex}}%
\stackon[1pt]{#1}{\tmpbox}%
}

\usepackage[table]{xcolor}
\newcommand{\blind}{0}

\usepackage{kantlipsum}
\usepackage[breakable, theorems, skins]{tcolorbox}
\tcbset{enhanced}

\usepackage{soul}
\usepackage{xcolor}     

\usepackage{calc}

\usepackage{tabularx}

\newcolumntype{b}{X}
\newcolumntype{s}{>{\hsize=.5\hsize}X}

\usepackage{booktabs}
\usepackage{caption}
\usepackage{geometry}

\newsavebox\CBox
\newcommand\hcancel[2][0.5pt]{%
  \ifmmode\sbox\CBox{$#2$}\else\sbox\CBox{#2}\fi%
  \makebox[0pt][l]{\usebox\CBox}%
  \rule[0.5\ht\CBox-#1/2]{\wd\CBox}{#1}}

\begin{document}

\bibliographystyle{agsm}

\def\spacingset#1{\renewcommand{\baselinestretch}%
{#1}\small\normalsize} \spacingset{1}


\if0\blind
{
  \title{\bf A fully Bayesian approach for the imputation and analysis of derived outcome variables with missingness}
  \author{Harlan Campbell$^{1,2}$, Tim P.\ Morris$^{3}$, Paul Gustafson$^{1}$ \\
    \\
  \small{1 -- Department of Statistics, University of British Columbia}\\
  \small{2 -- Health Economics and Outcomes Research, Precision AQ, Vancouver, BC, Canada}\\
  \small{3 -- MRC Clinical Trials Unit at UCL, University College London, London, UK}
  }
  \maketitle
} \fi

\if1\blind
{
  \bigskip
  \bigskip
  \bigskip
  \begin{center}
    {\LARGE\bf Title}
\end{center}
  \medskip
} \fi

\bigskip

\abstract{Derived variables are variables that are constructed from one or more source variables through established mathematical operations or algorithms.  For example, body mass index (BMI) is a derived variable constructed from two source variables: weight and height.  When using a derived variable as the outcome in a statistical model, complications arise when some of the source variables have missing values. In this paper, we propose how one can define a single fully Bayesian model to simultaneously impute missing values and sample from the posterior. We compare our proposed method with alternative approaches that rely on multiple imputation with  examples including an analysis to estimate the risk of microcephaly (a derived variable based on sex, gestational age and head circumference at birth) in newborns exposed to the ZIKA virus.}

\spacingset{1.45} 

\section{Introduction}

`Derived variables' are variables that are constructed from one or more measured variables through established mathematical operations or algorithms. Examples of commonly used derived variables include body mass index (BMI), constructed from weight and height, and the 36-Item Short Form Health Survey Physical Component Summary score (SF-36 PCS), constructed from a combination of responses measured on the 36-Item Short Form Survey (SF-36). 


When using a derived variable, a particular complication may arise when some of the measured `source' variables required to derive it have missing values. Missingness in source variables will often lead to missingness in derived variables.  For example, if weight is missing for certain study participants, their BMI cannot be derived.  However, this is not quite the same as having completely-missing information on the derived variable. In some instances, one might have some information.  For example, if a single question on the SF-36 is left unanswered, one might still be able to infer, at least approximately, a participant's SF-36 PCS based on their answers to the other 35 questions. More clearly, if a composite outcome is defined to be 1 if either component is 1, then observing that one component equals 1 means the composite is known whether or not the other component is measured \citep{Pham2021}.

The data that initially motivated this work consists of measurements taken from newborns of mothers infected with the ZIKA virus \citep{wilder2019understanding}.  In order to determine if the newborns have microcephaly (a derived variable), a calculation based on three source variables (sex, gestational age, and head circumference) is required.  The source variables are often subject to missingness and we were curious as to how one might be able to fit these data with an entirely Bayesian solution.

Besides so-called \textit{complete-case analysis}, in which observations with any missing values are discarded, the most popular statistical approach for handling missing data is multiple imputation (MI) \citep{rubin1976inference, reiter2007multiple} and among the different varieties of MI, the Multivariate Imputation by Chained Equations (MICE) approach is the most widely used \citep{van2018flexible}. The basic idea of MI is to fill in missing values by repeated simulation from the posterior predictive distributions of the missing values given assumptions about the missing data mechanism and a model, thereby preserving the uncertainty associated with the missingness.  MICE does this for multivariate non-monotone missing data variables by using a sequence of univariate conditional distributions (i.e., imputing variables one-at-a-time).  

When using MICE to impute missing values of derived variables, there are four general approaches:

\begin{enumerate}
    \item imputation at the derived-variable level (DVL), in which one imputes the derived variable directly ignoring the source variables;
    \item imputation at the source-variable level (SVL), in which one imputes the source variables first and then computes the derived variable afterwards (also known as the \textit{impute, then transform} approach); 
    \item the \textit{just another variable} (JAV), approach, in which one imputes the source variables and derived variable together, with the source variables as covariates in the imputation model for the derived variable.  To be clear, all variables are imputed separately, ignoring any deterministic relationships.  Instead, a multivariate normality between the derived variable and the sources variables is often assumed (e.g., \citet{von2009impute} and \citet{white2011multiple}).  Notably, \citet{seaman2012multiple} conclude that JAV should not be used to impute data for a logistic regression analysis as it performs ``very badly'' in this case; and
    \item  the \textit{``on-the-fly'' imputation}, where the transformation of the source variables into the derived variable is done ``on-the-fly'' within the imputation algorithm. (Note that \citet{van2018flexible} calls this ``passive imputation'').  The ``on-the-fly'' imputation consists of three steps. First, the source variables are imputed, then the derived variable is updated based on the known deterministic relationship with the source variables, and, third, all other variables are imputed conditional on this updated derived variable.  When missingness only occurs in the derived variable and source variables (i.e., no other variables need imputing) and missingness in source variables necessarily implies missingness in the derived variable (e.g., any individual with unknown height or weight will necessarily have unknown BMI), on-the-fly imputation is entirely equivalent to SVL imputation.
\end{enumerate} 

  The other approach we should mention, proposed by \citet{bartlett2015multiple}, is known as the Substantive Model Compatible Fully Conditional Specification  (SCMFCS) imputation.  This method is useful for imputing derived covariate values, but not derived outcome variables (which is our focus).  

There may be challenges when imputing at the source-variable level if the derived variable is not a linear function of the source variables; see \citet{seaman2012multiple}, \cite{morris2014multiple}, and \citet{tilling2016appropriate}. However, because DVL imputation effectively excludes information available in the source variables (which are possibly the best predictors of the missing derived variable values), SVL (or ``on-the-fly'') imputation is thought to be a more efficient strategy, in the sense that it uses observed information where possible.  \citet{pan2020passive} compare the performance of the DVL and SVL approaches and conclude that SVL imputation is indeed preferable, regardless of how the derived variable is obtained (``whether by simple arithmetic operations or by some highly specific algorithms'').  This follows related work by \citet{gottschall2012comparison} who arrived at a similar conclusion for handling missing values in questionnaire data (``researchers should adopt item-level imputation whenever possible'').

\begin{table}
\begin{center}
\begin{tabular}{l|cccc}
    \hline
   & \(Z_1\) (source) & \(Z_2\) (source) & & \(Y\) (derived) \\ \hline
    Pattern 1 & Missing  & Observed & $\to$ & Missing  \\
    Pattern 2 & Observed & Missing & $\to$ & Missing  \\
    Pattern 3 & Missing  & Missing & $\to$ & Missing  \\ \hline
    \end{tabular}
\caption{Three patterns of missingness in \(Z_1,Z_2\) that result in \(Y\) being missing, where \(Y=Z_1 + Z_2\). \label{tab:patterns}}
\end{center}
\end{table}    

Another reason to favour the SVL, ``on-the-fly'', and JAV approaches  over DVL imputation is that they make a less restrictive MAR (missing at random) assumption. MAR requires that the conditional probability of missingness depends only on data observed under that pattern \citep{rubin1976inference}. Consider for example a derived outcome \(Y\) that is equal to the sum of two source variables, $Z_1$ and $Z_2$, such that $Y=Z_1 +Z_2$. Missingness in \(Y\) can occur in three patterns:  First, the $Z_1$ value is missing, second the $Z_2$ value is missing, third, both $Z_1$ and $Z_2$ values are missing; see Table~\ref{tab:patterns}. Because MAR is i) defined at the \textit{observation} level rather than at the variable level, and ii) conditional on variables included in the analysis, we see that assuming MAR with DVL imputation requires that missingness in the derived variable \(Y\) be independent of the value of \(Z_1,Z_2\) under any pattern, see Table \ref{tab:patterns2}. For the SVL, ``on-the-fly'', and JAV approaches, this is relaxed under patterns 1 and 2: missingness in \(Z_1\) under pattern~1 need only be independent of \(Z_1\), not \(Z_2\), and missingness in \(Z_2\) under pattern~2 need only be independent of \(Z_2\), not \(Z_1\).

\begin{table}
\begin{center}
    \begin{tabular}{l|ccc}
    \hline
    & \(A=Z_1\) (source) & \(A=Z_2\) (source) & \(A=Y\) (derived) \\ \hline
    &\multicolumn{3}{c}{The MAR assumption requires that, under this} \\
    &\multicolumn{3}{c}{pattern, missingness in $A$ is independent of} \\ \hline
    Pattern 1: & \(Z_1\) & - & \(Z_1, Z_2\) \\
    Pattern 2: & - & \(Z_2\) & \(Z_1, Z_2\) \\
    Pattern 3: & \(Z_1, Z_2\) & \(Z_1, Z_2\) & \(Z_1, Z_2\) \\ \hline
\end{tabular}
\caption{For each of the patterns in table~\ref{tab:patterns}, this table shows the what the MAR assumption requires missingness be independent of. For the DVL imputation approach, missingness in the \(Y\) must be independent of the value of $Z_1$ and the value of $Z_2$, regardless of pattern.\label{tab:patterns2}}
\end{center}
\end{table}

To conduct Bayesian inference with imputed data, \citet{gelman2004bayesian} outline a three step procedure. The first step is to construct $K$ imputed datasets, using MI. The second step is to simulate draws from the posterior distribution of the model parameters with each imputed dataset separately.  The third step is to mix all the draws together, thereby obtaining (at least approximately) draws from the posterior distribution that take into account the uncertainty associated with the missing data; see \citet{zhou2010note}. Summaries (e.g., median, 2.5th, and 97.5th percentiles) of the mix of draws from across the imputed datasets are then obtained for one's parameter estimates.  (This third step is perhaps surprising to those familiar with multiple imputation inference: Rubin's combining rules are not needed.)  While \citet{gelman2004bayesian}'s three-step procedure appears straightforward, when SVL (or ``on-the-fly'') imputation is used for imputing a derived \textit{outcome} variable, things can be much more complicated.  This becomes apparent when considering the assumptions required in the first two steps.  

Using SVL or ``on-the-fly'' imputation in the first step requires one to specify distributions for each of the missing source variables.  Then, specifying the posterior distribution in second step requires one to define a distribution -- along with priors for the parameters that characterize the distribution -- for the derived outcome variable.  Clearly the assumed distributions of the source variables (specified in step~1) necessarily determine the distribution of the derived outcome variable (which must be specified in step~2).  Without careful consideration, one could inadvertently define incompatible distributions, since the two steps are done separately and the software implementing one step (e.g., MICE from \citet{van2015package}) does not know of the assumptions required by the other step (e.g., rstan from \citet{carpenter2017stan}).  Moreover, it will no doubt be difficult to determine whether or not the assumptions made by the Bayesian model specified in the second step (e.g., regarding priors) are consistent with the assumptions made when imputing missing values in the first step.  

Ideally, one could fit a single fully Bayesian model to simultaneously impute missing values and sample from the posterior. The idea of avoiding potential problems with MI by adopting a fully Bayesian approach is not new and it is actually quite common to proceed in this way when covariate values are missing; see \citet{bartlett2015multiple}, \citet{ludtke2020regression} and \citet{Erler2021} and the references therein. However, it is not at all obvious how to proceed when there is missingness in source variables that combine to define a derived outcome variable. In this article, we propose a useful approach.

In Section 2, we begin with a simple illustrative example.  In Section \ref{sec:description}, we outline our proposed approach. In Section 4, we consider two applied examples: (1) comparing the BMI of Dutch boys from inside the city to those from outside the city, and (2) estimating the risk of microcephaly in newborns exposed to the ZIKA virus with an artificial dataset. Finally, we conclude in Section 5.

\section{A simple illustrative example}

We continue with the simple example discussed in the introduction where $Y=Z_1 +Z_2$.  Suppose we are interested in two groups, group $A$ and group $B$, (e.g., placebo and treatment) and we are interested in estimating  the difference between the mean of $Y$ in group $A$, and the mean of $Y$ in group $B$.   The easiest way to proceed might be to fit a simple univariate Normal model for $Y$ where, for $i = 1,\ldots,n$:
\begin{align}
y_{i} &\sim \text{Normal}(\alpha + \beta x_{i}, \sigma_{y}^{2}) \quad \label{eq:univariate} \end{align}
where $x_{i} = 0$ if observation $i$ is in group $A$, and $x_{i} = 1$ if observation $i$ is in group $B$,
with priors specified for $\alpha$, $\beta$ and $\sigma_{y}$.  The estimand (the target quantity that is to be estimated) coincides with $\beta$.

Now suppose that, in group $A$, a proportion of $Z_1$ and $Z_2$ values are missing.  Specifically, in group $A$, certain observations are prone to have missing $Z_1$ values with the chance of missingness depending on $Z_2$, and other observations are prone to have missing $Z_2$ values with the chance of missingness depending on the observed $Z_1$.  (To be clear, in this specific example, no observations are susceptible to missing both $Z_1$ and $Z_2$.)  Table \ref{tab:data1} lists the data for six individuals to illustrate, with ``NA'' indicating missing. Given this missingness in the data, how should we proceed to estimate $\beta$?   

\begin{table}[ht]
\centering
\begin{tabular}{rlrrr}
  \hline
  $i$ & X (Group) & $Y$ & $Z_{1}$ & $Z_{2}$ \\ \hline
  1 & A & NA & NA & 1.71 \\
  2 & A & 1.59 & $-$0.08 & 1.68 \\ 
  3 & A & NA & 1.23 & NA \\ 
  4 & B & 2.40 & 2.20 & 0.20 \\ 
  5 & B & 4.49 & 2.35 & 2.14 \\ 
  6 & B & 3.53 & 0.49 & 3.03 \\ \hline
\end{tabular}
\caption{Six observations from the hypothetical dataset considered in the example.}
\label{tab:data1}
\end{table}

As discussed in the introduction, one could consider the complete-case analysis, where all rows of data with any missing values are  discarded (individuals $i=1$ and $i=3$ listed in Table \ref{tab:data1}).  The advantage of this strategy is that no imputation step is required and the entire procedure is to simply fit the univariate Normal model. The disadvantage is that this strategy almost certainly involves a loss of efficiency and has the potential for bias. 

Alternatively, one could apply Gelman's three step procedure  with one of the three MICE approaches (either the DVL, SVL, or JAV imputation methods) for imputing the missing $Y$ values followed by the univariate Normal model for inference.  The advantage of this strategy is that, depending on the specific MICE approach used, one could conceivably obtain unbiased estimates of $\beta$ in an efficient manner.  The disadvantage is that this strategy involves two separate steps (an imputation step and an inference step) which may not be consistent with one another.

There is another possible strategy worth considering.  Because the math in this scenario is rather simple, one could fit a Bayesian model for the source variables, $Z_1$ and $Z_2$, and derive the ``implied'' samples of $\beta$.    For instance, we could fit a bivariate Normal model for $Z_{1}$ and $Z_{2}$, where, for $i$ in 1,...,$n$:
\begin{align}
 (Z_{1,i}, Z_{2,i}) &\sim \text{Normal}(\boldsymbol{\mu_{i}}, 
\boldsymbol{\Sigma}),
\label{eq:bivariateNormal}
\end{align}
where
\begin{align}
\boldsymbol{\mu_{i}} &= \begin{cases} 
(\mu_{Z1,A}, \mu_{Z2,A}) & \text{if $i$ is in group $A$} \nonumber \\ 
(\mu_{Z1,B}, \mu_{Z2,B}) & \text{if $i$ is in group $B$}  \nonumber \\
\end{cases}\\
\textrm{and} \quad \quad
\boldsymbol{\Sigma} &=
\begin{pmatrix}
\sigma_{Z1}^2 & \rho\sigma_{Z1}\sigma_{Z2} \\
\rho\sigma_{Z1}\sigma_{Z2} & \sigma_{Z2}^2 \\
\end{pmatrix}, \nonumber \\
\textrm{\noindent and, for each MCMC iteration,}&\textrm{ compute} \quad \quad \nonumber \\
\beta &= (\mu_{Z1,A} + \mu_{Z2,A}) - (\mu_{Z1,B} + \mu_{Z2,B}). \label{eq:easymath}
\end{align}
Priors would need to be specified for each of the seven parameters: $\mu_{Z1,A}$, $\mu_{Z2,A}$, $\mu_{Z1,B}$, $\mu_{Z2,B}$ $\sigma_{Z1}$, $\sigma_{Z2}$ and $\rho$.

This strategy might be ideal in the sense that it does not require a separate imputation step (the entire procedure is to simply fit the bivariate Normal model) and does not disregard a large number of observations.  However, it is possible only because in this simple illustrative example the math is extremely straightforward (i.e., equation (\ref{eq:easymath}) is known).  For other derived variables, this might not be the case.  For instance, if instead of $Y=Z_{1} + Z_{2}$, we had $Y=Z_{1}/({Z_{2}}^{2})$, then computing $\beta$ from a combination of the seven model parameters ($\mu_{Z1,A}$, $\mu_{Z2,A}$, $\mu_{Z1,B}$, $\mu_{Z2,B}$ $\sigma_{Z1}$, $\sigma_{Z2}$ and $\rho$) might be impossible without a degree of mathematical wizardry.

This brings us to our proposed method, details of which are provided in the next section.  Briefly, the approach involves using Monte Carlo integration to derive samples of the derived outcome variable. In our example this would involve fitting the bivariate Normal model and sampling $\beta$ (at least approximately) without requiring any knowledge about the mathematical derivation of $\beta$.

\section{Description of the approach}
\label{sec:description}
Having established the motivation behind our proposed method, we now outline the details.

Suppose the data consists of $n$ observations and let $W$ be the set of $p+q$ variables at hand.  Let $Y = f(W) = f(W_{source})$ be the outcome variable, where $W_{source}$ are the $p$ source variables that combine in some deterministic way to equal the outcome variable. (To be clear, there are $p$ source variables which make up the outcome $Y$). Let $W_{expl}$ consist of the remaining $q$ explanatory variables.  \textcolor{black}{In practice, note that certain variables could be both source variables and explanatory, but we set this aside to avoid more complicated notation}.  Here,  $\dim(W_{source})=(n,p)$ and $\dim(W_{expl})=(n,q)$. 

Say $W$ partitions as variables $W_{1}$ which are prone to missingness, and variables $W_{2}$ which are not:  $W = \{W_{1}, W_{2}\}$. Specifically, $W_{1}$ consists of certain observations that are missing, $W_{1}^{\{mis\}}$, and certain observations that are observed, $W_{1}^{\{obs\}}$, such that $W_{1}=\{W_{1}^{\{mis\}}, W_{1}^{\{obs\}}\}$.  We are interested in the situation in which there is overlap between $W_{1}$ and $W_{source}$, i.e.,  situations when certain source variables are subject to missingness.  Let $W_{1,source} = W_{1} \cap W_{source}$, $W_{1,expl} = W_{1} \cap W_{expl}$, $W_{2,source} = W_{2} \cap W_{source}$, and $W_{2,expl} = W_{2} \cap W_{expl}$.

In the simple illustrative example we have $W = \{Z_{1}, Z_{2}, X\}$ (where $X$ is the binary group indicator variable such that, for the $i$-th observation, $X_{i}=0$ if $i$-th observation is in Group A and $X_{i}=1$ if the $i$-th observation is in Group B). Furthermore, we have $W_{1} = \{Z_{1}, Z_{2}\}$, $W_{2} = \{X\}$, $p=2$, $W_{source} = \{Z_{1}, Z_{2}\}$, $q=1$, $W_{expl}=\{X\}$, and $Y = f(W) = Z_{1} + Z_{2}$.  

Typically, the objective of a Bayesian analysis consists of  the estimation of an estimand \citep{kahan2023eliminating}.  The estimand of interest, $\theta$, is often  defined non-parametrically as a function of certain conditional expectations: $\theta = g(\textrm{E}^{*}_1 \{ \textrm{E}(Y|W_{expl}) \},..., \textrm{E}^{*}_D \{ \textrm{E}(Y|W_{expl}) \})$, where $\textrm{E}^{*}_d$ corresponds to the counterfactual expectation with respect to  the $d$-th population of interest for $d$ in 1,...,$D$. 

In certain situations, the target population(s) might themselves be unknown and require estimation. For example, suppose the estimand of interest is the mean difference in $Y$ between those receiving treatment ($X=1$) and those not receiving treatment ($X=0$) for individuals who are of ``average age'', where average age, $\mu_{A}$, is unknown.  Then, we have $D=2$ target populations: (1) treated individuals of average age, and (2) untreated individuals of average age. The estimand is defined as: $\theta = \textrm{E}(Y|X=1,A=\mu_{A}) - \textrm{E}(Y|X=0,A=\mu_{A})$.  The parameter $\mu_{A}$ is required to define the two target populations and must be estimated from the data.

In the simple illustrative example, the estimand of interest is the difference in the mean of $Y$ between individuals in the two groups (i.e., the difference between the conditional expectation of $Y$ given the $D=2$ target populations defined by $X=0$ and $X=1$): $\theta = \textrm{E}(Y|X=1) - \textrm{E}(Y|X=0)$.  Because the math is conveniently simple, one can also define this estimand analytically as a function of the model parameters: $\theta = \beta$, but also as: $\theta = h(\phi_{source}) = (\mu_{Z1,A} + \mu_{Z2,A}) - (\mu_{Z1,B} + \mu_{Z2,B})$, where $\phi_{source} = \{\mu_{Z1,A}$, $\mu_{Z2,A}$, $\mu_{Z1,B}$, $\mu_{Z2,B}$ $\sigma_{Z1}$, $\sigma_{Z2}$, $\rho\}$. The standard Bayesian approach would then be to obtain a Monte Carlo sample of size $M$  (e.g., via MCMC) from the $(\phi_{source},W_{1}^{\{mis\}}|W_{1}^{\{obs\}},W_{2})$ posterior distribution and then, since $h()$ is known, obtain $M$ posterior samples of $\theta$ by applying the $h()$ function to each of the $M$ posterior samples of $\phi_{source}$.
 
 In situations when the math is not so simple, Bayesian g-computation \citep{gustaf2015,keil2018bayesian} can be used for a form of `model-based standardization' \citep{zou2009assessment, vansteelandt2011invited} in order to obtain posterior samples of $\theta$ as follows.  First, for each of the target populations of interest, one forward samples  ``posterior predictive values'' of the source variables conditional on the target population of interest. Then, by applying the deterministic function $f()$ on the sample of these ``posterior predictive values'' one can obtain a  sample of $Y^{*}_{d}$, a new variable with the same identical distribution as $Y$ under the $d$-th population of interest, i.e., a ``future observation'' projected by the model of the $Y$ variable from the $d$-th target population.  Finally, with a large number, $S$, of these ``future observations'' one can approximate $\textrm{E}^{*}_{d}(\textrm{E}(Y|W_{expl}))$, for $d$ in 1,...$D$, and combine these according to the definition of the estimand of interest to obtain a posterior sample of $\theta$.  Obtaining samples of $\theta$ in this way ensures that the correct amount of variation is captured and, importantly, does not require knowledge of a function $h: \phi_{source} \rightarrow \theta$.  
 
To be clear this approach consists of the following steps:

\begin{enumerate}
\item{Draw $M$ samples of $(\phi_{source}, W_{1,expl}^{\{mis\}})$ from the $(\phi_{source}, W_{1,expl}^{\{mis\}}|W_{1}^{\{obs\}},W_{2})$ distribution: $(\phi_{source}, W_{1,expl}^{\{mis\}})^{[m]}$, for $m$ in 1,...,$M$.}
 \item{For $m$ in 1,...,$M$:
 \begin{enumerate}
 \item{For $d$ in 1,...,$D$:
 \begin{enumerate}
    \item If the $d$-th target population is defined in terms of unknown parameters, $\gamma^{d}$, these must be estimated from the data.  Obtain a draw from $\gamma^{d}|(\phi_{source}, W_{1,expl}^{\{mis\}})^{[m]}$:  $\gamma^{d[m]}$.
     \item {Draw $S$ (where $S$ is a large number) iid samples from the $(W_{source}|\gamma^{d[m]})$:  $W^{*d[m,s]}_{source}$, for $s$ in 1,..., $S$.  Note: $\dim(W^{*d[m,s]}_{source}) = (1,P)$.}
     \item{For $s$ in 1,...,$S$: Apply the deterministic function $f()$ to $W^{*d[m,s]}_{source}$ to obtain a sample from the $(Y|\gamma^{d[m]})$ distribution:   $Y^{*d[m,s]} = f(W^{*d[m,s]}_{source})$.  Note: $\dim(Y^{*d[m,s]}) = (1,1)$.}
     \item{Calculating (approximately) the sample mean: \newline \mbox{ $\textrm{E}^{*}_d \{ \textrm{E}(Y|W_{expl}) \}^{[m]} \approx \frac{1}{S}\sum_{s=1}^{S}{Y^{*d[m,s]}}$}.}
     \end{enumerate}
     }
    \item{Obtain ${\theta}^{[m]}$ by combining $\textrm{E}^{*}_1 \{ \textrm{E}(Y|W_{expl}) \},..., \textrm{E}^{*}_D \{ \textrm{E}(Y|W_{expl}) \}$ according to the definition of the estimand of interest.}
     
 \end{enumerate}
     }  

 \end{enumerate}

Having obtained $M$ posterior samples of $\theta$ following the four steps, one can calculate the posterior median/mean and credible interval for $\theta$, as is standard practice in a Bayesian analysis.  Note that both $M$ and $S$ should be sufficiently large that \textit{both} sources of Monte Carlo error are negligible. 

In the simple illustrative example, the $D=2$ target populations were defined by $X=0$ and $X=1$.  Therefore the $m$-th posterior sample of $\theta$ is approximated as:
\begin{equation}
\theta^{[m]} = \frac{1}{S}\sum_{s=1}^{S}{Y^{*2[m,s]}}  - \frac{1}{S}\sum_{s=1}^{S}{Y^{*1[m,s]}},
\end{equation}
where $Y^{*1[m,s]}$ is a sample from the distribution of  $(Y|\phi_{source}^{[m]}, X=0)$ and $Y^{*2[m,s]}$ is a sample from the distribution of $(Y|\phi_{source}^{[m]}, X=1)$.

\section{Applied examples}

\subsection{Multiple linear regression of the body mass index of Dutch boys}

\subsubsection*{Background}

The ``Growth of Dutch boys'' dataset is available as an example dataset within the `mice' package for R \citep{van2015package} and is often used to demonstrate how various imputation methods can be applied.  Here we use this dataset to illustrate how the approach outlined in Section \ref{sec:description} could be applied for determining the difference in logBMI between boys from the ``city'' and boys from outside of the ``city'', when controlling for age.

The outcome of interest, $Y=\textrm{log}(\textrm{BMI})$, is a derived binary variable, based on two source variables: (1) height ($Z_{1}$, measured in centimeters), (2) weight ($Z_{2}$, measured in kilograms). This is derived as follows:
\begin{equation}
    Y = \textrm{logBMI} = f(Z_{1},Z_{2})= \textrm{log}\Big(\frac{Z_{2}}{({Z_{1}/100})^{2}}\Big).
\end{equation}
The exposure variable is regional status ($X$, equal to 1, if the boy is from the ``city'' and equal to 0 otherwise).  We also consider the variable age ($A$, measured in years) as highly predictive of $Z_{1}$, $Z_{2}$, and $Y$.

\subsubsection*{The data}

\begin{figure}
    \centering
    \includegraphics[width=0.8\linewidth]{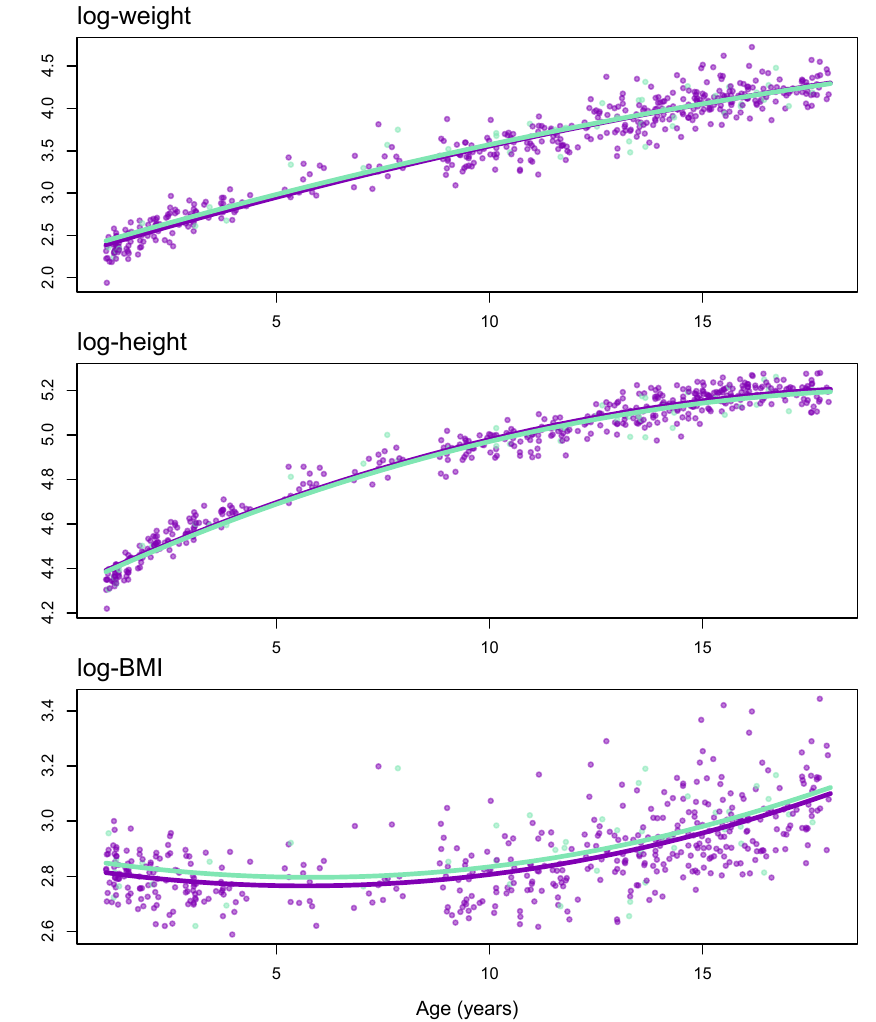}
    \caption{The complete case data for the Dutch boys analysis ($n$=518) with violet dots corresponding to boys from outside the city ($X=0$) and green dots corresponding to boys from the city ($X=1)$.  Lines drawn from least-squares estimates.}
    \label{fig:data_boys}
\end{figure}

\begin{figure}
    \centering
    \includegraphics[width=0.8\linewidth]{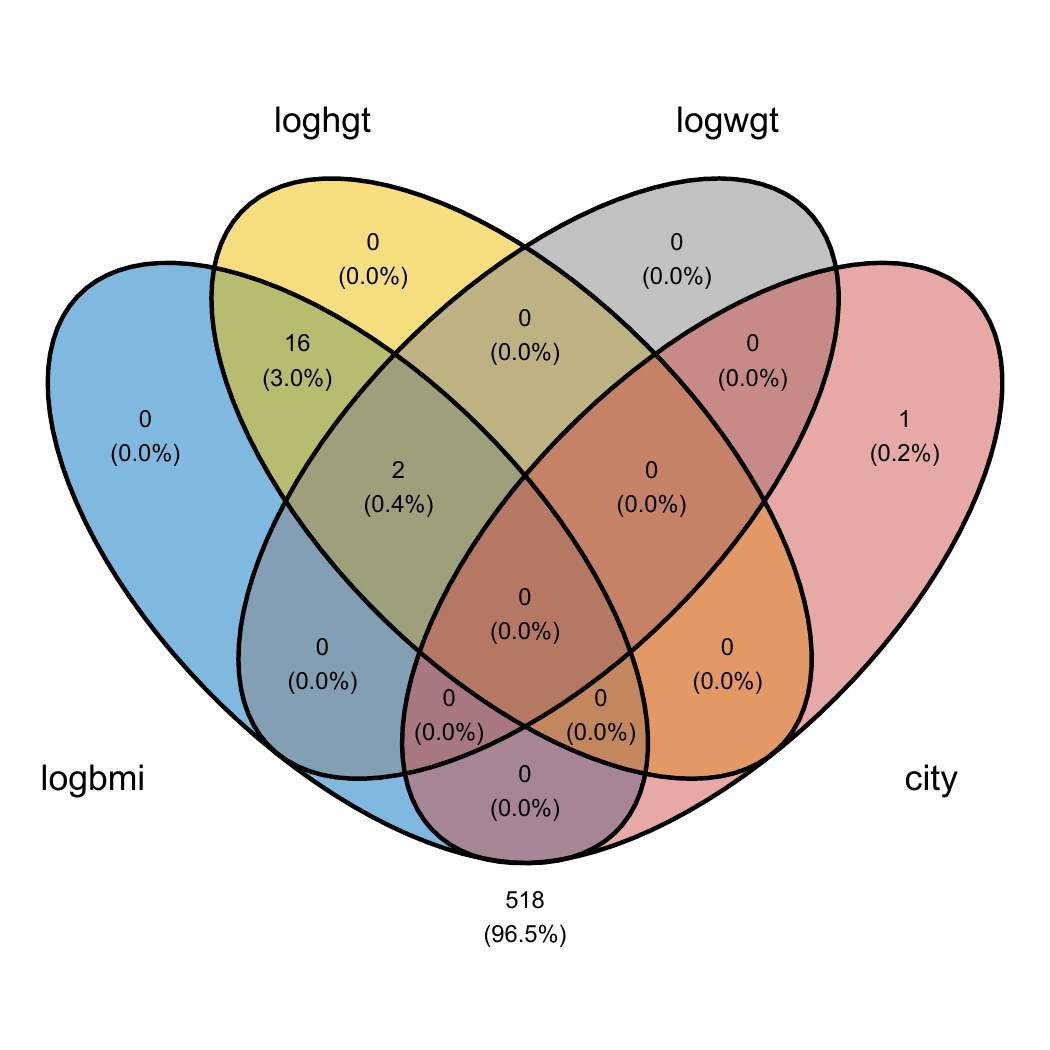}
    \caption{Venn diagram for missingness in Dutch boys dataset ($n$=537).}
    \label{fig:venn_boys}
\end{figure}

The data consists of measurements on a representative sample of $n=537$ boys between the ages of 1 and 18 years (inclusive), of which approximately 9\% are from the ``city'' (we restricted the original ``Growth of Dutch boys'' dataset dataset ($n=748$) to the 1--18 age range to allow for reasonable modeling of how \textrm{log}(\textrm{BMI}) changes with age).  Figure \ref{fig:data_boys} plots the data among the complete cases.  All variables, except age ($A$), are subject to a small amount of  missingness.  The exposure variable, ``city'' ($X$), is missing for 1 subject; height is missing for 18 subjects; and weight is missing for 2 subjects; see Venn diagram in Figure \ref{fig:venn_boys}.   Overall, 19 out of the 537 of observations (3.5\%) have at least one missing value.  Following the notation defined in Section \ref{sec:description}, we have that $W_{1}=\{Z_{1}, Z_{2}, X\}$, $W_{2}=\{A\}$, $p=2$, $W_{source}=\{Z_{1}, Z_{2}\}$, $q=2$, and $W_{expl}=\{X, A\}$.  

The primary objective of the study is to estimate the difference in mean logBMI between boys from inside the ``city'' and boys from outside of the ``city'', when controlling for age.  A secondary objective is to make probabilistic predictions of logBMI for boys given their age and regional status.

\subsubsection*{The estimand}

Based on the primary objective, the estimand of interest is the difference in mean logBMI between boys from the ``city'' and boys from outside of the ``city'', when controlling for age, and this can be defined as a function of the conditional expectation of $Y$ given $D=2$ target populations:
\begin{align}
\theta &= \int_{a}\textrm{E}(Y|X=1, A=a)f_{A}(a)da - \int_{a}\textrm{E}(Y|X=0, A=a)f_{A}(a)da,
\end{align}
where $f_{A}(a)$ corresponds to the distribution of ages among boys in the Netherlands.  Since the data supposedly consist of a representative sample of Dutch boys, we assume that $f_{A}(a)$ also coincides with the sample distribution of ages.

\subsubsection*{Two models}

The univariate normal model involves considering the logBMI of each subject as the outcome of a simple univariate Normal model:
\begin{equation}
    (Y|X=x, A=a) \sim \textrm{Normal}(\beta_{0} + \beta_{1}x + \beta_{2}a + \beta_{3}xa + \beta_{4}a^{2}, \sigma^{2}).
\end{equation}
This model assumes a quadratic relationship between BMI and age, a reasonable approximation based on published growth curves  \citep{kuczmarski2000cdc} and allows for the possibility that age is an effect modifier with respect to the effect of ``city''. For the univariate normal model, we set standard Normal priors, $\textrm{Normal}(0,1)$, for the regression coefficient parameters $\beta_{0}$, $\beta_{1}$, $\beta_{2}$, $\beta_{3}$ and $\beta_{4}$, an exponential prior, $\textrm{Exponential}(1)$, for the standard deviation parameter, $\sigma$, and a uniform prior, $\textrm{Uniform}[0,1]$, for $\pi$.

The bivariate normal model considers height and weight as two correlated outcomes such that:
\begin{align}
 (\textrm{log}(Z_{1}), \textrm{log}(Z_{2})|X=x, A=a) &\sim \text{Normal}(\boldsymbol{\mu_{i}}, 
\boldsymbol{\Sigma}),
\label{eq:bivariateNormal}
\end{align}
where
\begin{align}
\boldsymbol{\mu_{i}} &=  
\begin{pmatrix}\alpha_{0} + \alpha_{1}r + \alpha_{2}a + \alpha_{3}xa + \alpha_{4}a^{2}\\
\gamma_{0} + \gamma_{1}x + \gamma_{2}a + \gamma_{3}xa + \gamma_{4}a^{2}
\end{pmatrix}
\textrm{and} \quad \quad \\
\boldsymbol{\Sigma} &=
\begin{pmatrix}
\tau_{Z1}^2 & \rho\tau_{Z1}\tau_{Z2} \\
\rho\tau_{Z1}\tau_{Z2} & \tau_{Z2}^2 \\
\end{pmatrix}.\nonumber 
\end{align}
and:
\begin{equation}
    R \sim \textrm{Bernoulli}(\pi).
\end{equation}
We set standard Normal priors, $\textrm{Normal}(0,1)$, for all the regression coefficient parameters $\alpha_{0}$, $\alpha_{1}$, $\alpha_{2}$, $\alpha_{3}$, $\alpha_{4}$, $\gamma_{0}$, $\gamma_{1}$, $\gamma_{2}$, $\gamma_{3}$, and $\gamma_{4}$; exponential priors, $\textrm{Exponential}(1)$, for the standard deviation parameters, $\tau_{Z1}$ and $\tau_{Z2}$, a uniform prior, $\textrm{Uniform}[-1,1]$, for the correlation parameter, $\rho$, and another uniform prior, $\textrm{Uniform}[0,1]$, for $\pi$.

We can define the estimand of interest analytically as a function of the parameters of the univariate normal model:
\begin{align}
\theta = h_{1}(\phi_{univ}) &= \int_{a}{(\beta_{1} + \beta_{3}a)f_{A}(a)}da\\
&= \beta_{1} + \beta_{3}\textrm{E}(A),
\end{align}
where $\phi_{univ}= \{\beta_{0}, \beta_{1}, \beta_{2}, \beta_{3}, , \beta_{4}, \sigma\}$.  Since the data is assumed to be a representative sample from the target population, and since we are choosing not to model the distribution of age (not required because there are no missing values for age), we can substitute $\textrm{E}(A)$ for the sample mean, $\bar{A}$.

Coincidentally, we can define the estimand as a function of the parameters of the bivariate normal model (see details in the Appendix):
%
\begin{align}
\theta = h_{2}(\phi_{biv}) &= \int_{a}{(\gamma_{1} + \gamma_{3}a -2\alpha_{1} - 2\alpha_{3}a)f_{A}(a)}da \nonumber \\ 
&=\gamma_{1} -2\alpha_{1} + (\gamma_{3} - 2\alpha_{3})\textrm{E}(A),
\end{align}
where $\phi_{biv}=\{\alpha_{0}, \alpha_{1}, \alpha_{2},  \alpha_{3}, \alpha_{4}, \gamma_{0}, \gamma_{1}, \gamma_{2}, \gamma_{3},  \gamma_{4},  \tau_{Z1}, \tau_{Z2}, \rho\}$, and we once again can substitute  $\textrm{E}(A)=\bar{A}$.

\subsubsection*{Methods and Results}

We conducted the analysis with four different approaches:

\begin{enumerate}
    \item  Fitting the univariate normal model to the complete case data and applying the $h_{1}()$ function to obtain posterior draws of $\theta$ (with $M=2,000$).
    \item Fitting the univariate normal model with Gelman's three-step approach with ``on-the-fly'' multiple imputation (including a quadratic term for age and age-city interaction term in the imputation model), and applying the $h_{1}()$ function to obtain posterior draws of $\theta$ (with $M=2,000$, and $K=50$ imputed datasets).
    \item Fitting the bivariate normal model and applying $h_{2}()$ function to obtain posterior draws of $\theta$ (with $M=2,000$).
    \item  Fitting the bivariate normal model and applying the proposed method detailed in Section \ref{sec:description} (with $M=2,000$ , and $S=2,000$).
\end{enumerate}

Both the univariate model and bivariate model were fit using JAGS with 1,000 burn-ins \citep{plummer2004jags}.  R and JAGS code is available in the Appendix.

Results are listed in Figure \ref{fig:boys_results}.  The estimate from the complete case analysis is similar to the estimates obtained when using the the bivariate model with a wider credible interval, suggesting that one can gain precision by incorporating the small number of observations with missing values into the analysis.  However, one obtains a wider credible interval when using the  univariate normal model with Gelman's three-step approach with ``on-the-fly'' multiple imputation, suggesting that precision can also be lost.  Finally, the estimate obtained with fitting the bivariate normal model and applying $h_{2}()$ function is slightly more precise than the estimate obtained with fitting the bivariate normal model and applying the proposed method, suggesting that there is a small cost to the numerical approximation (even with the large $S=2,000$).  With respect to computational time, the proposed method is substantially more time consuming than the other methods.

\begin{figure}
    \centering
    \includegraphics[width=\linewidth]{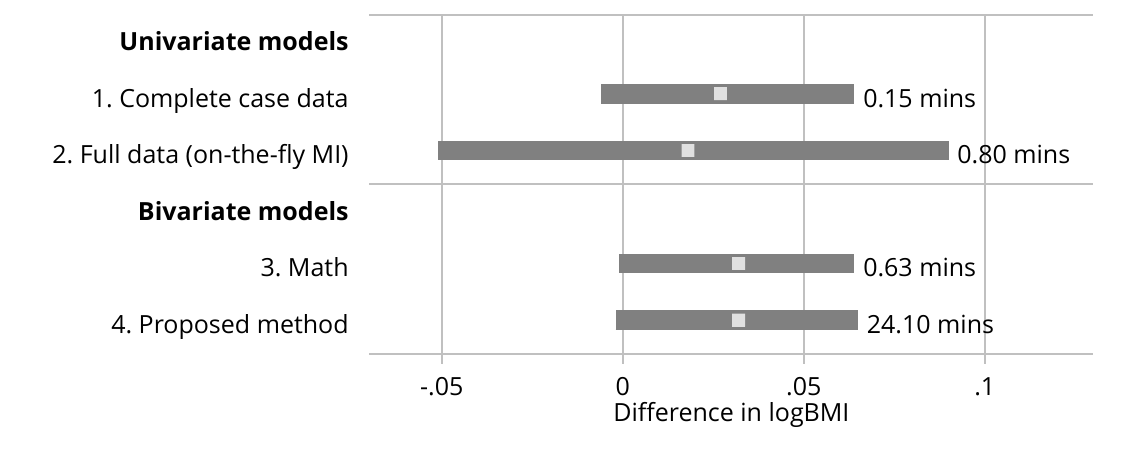}
    \caption{Results for the analyses of the Dutch boys dataset.  Estimates list the posterior median estimates (i.e., the median of the posterior estimates of the mean difference in logBMI) and equal-tailed posterior 95\%CrIs. Labels indicate computational time required for each method in minutes.}
    \label{fig:boys_results}
\end{figure}

\subsection{Estimating the risk of microcephaly}

\subsubsection*{Background}

Zika virus infection (ZIKV) during pregnancy is known to be associated with an increased risk of fetal congenital malformations, including microcephaly, a birth condition in which a baby's head is smaller than expected when compared to babies of the same sex and age.  Formally, microcephaly is defined as a measure of head circumference at birth of more than 2 standard deviations below average for sex and gestational age; see \citet{harville2020measurement}. (Average values for head circumference at birth are determined by referencing the standards calculated by the \textsc{intergrowth}-21st research network; see \citet{villar2014international}).  Therefore, in a healthy population, we would anticipate 2.28\% of newborns being categorized as having microcephaly (assuming head circumference in the general population is Gaussian).

In order to determine the risk of microcephaly associated with ZIKV in mothers, observational data has been collected by numerous studies from infected pregnant women \citep{zika2020zika}.  These data typically suffer from substantial missingness.

The outcome of interest, microcephaly ($Y$), is a derived binary variable, based on three source variables: (1) sex ($Z_{1}$=0 for `male' and $Z_{1}$=1 for `female'), (2) gestational age ($Z_{2}$, measured in weeks), and (3) head circumference ($Z_{3}$, measured in centimetres).  Each of these three source variables is subject to missingness.  The estimand of interest is the probability of microcephaly ($\theta = \textrm{E}(Y)$) given then assumed $D=1$ target population from which the data is sampled.


\subsubsection*{The data and models}

\begin{figure}
    \centering
    \includegraphics[width=15cm]{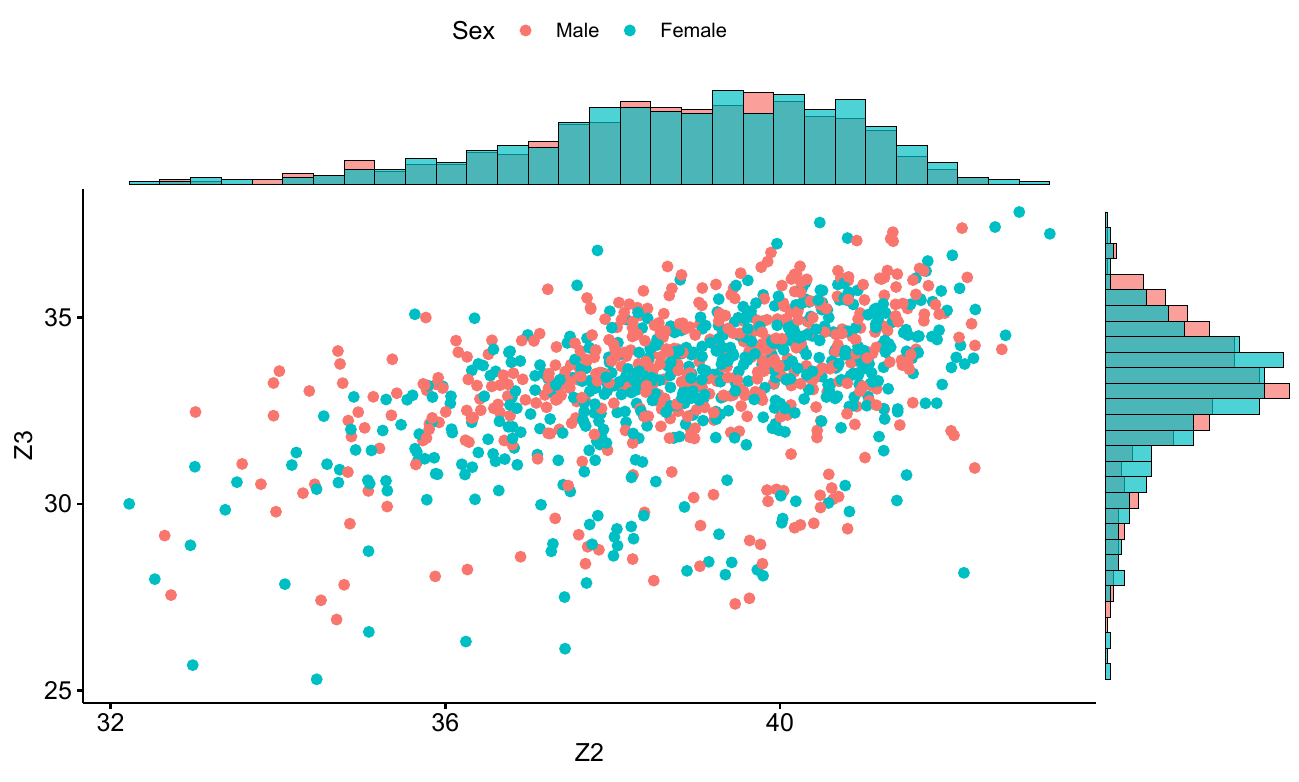}
    \caption{The complete case data ($n$=1,019) with values (1) sex ($Z_{1}$=0 for `male' and $Z_{1}$=1 for `female'), (2) gestational age ($Z_{2}$, measured in weeks), and (3) head circumference ($Z_{3}$, measured in centimetres). }
    \label{fig:dotplot}
\end{figure}

We consider a simple synthetic dataset based on what is typically observed in observational studies.  The purpose of this analysis is to demonstrate all of the steps required when following each of the different approaches for dealing with missingness based on the problem that initially motivated our research.  As such, we stress that the specific details of how the data were created are less important and acknowledge that a different dataset would no doubt lead to different results.  In brief, the data is simulated with a ``true'' risk of microcephaly of $\theta$ = 11.69\%.  The data represent $Z_1,Z_2,Z_3$ measurements on $n=1,800$ newborns with at least one of $Z_1,Z_2,Z_3$ observed; Figure~\ref{fig:dotplot} shows $Z_3$ and $Z_2$ among those with both observed.  Missing $Z_{1}$ values are attributed completely at random, missing $Z_{2}$ values are more likely for those with very small values of $Z_{3}$, and missing $Z_{3}$ values are more likely for those with very large values of $Z_{2}$.

\begin{figure}
    \centering
    \includegraphics[width=15cm]{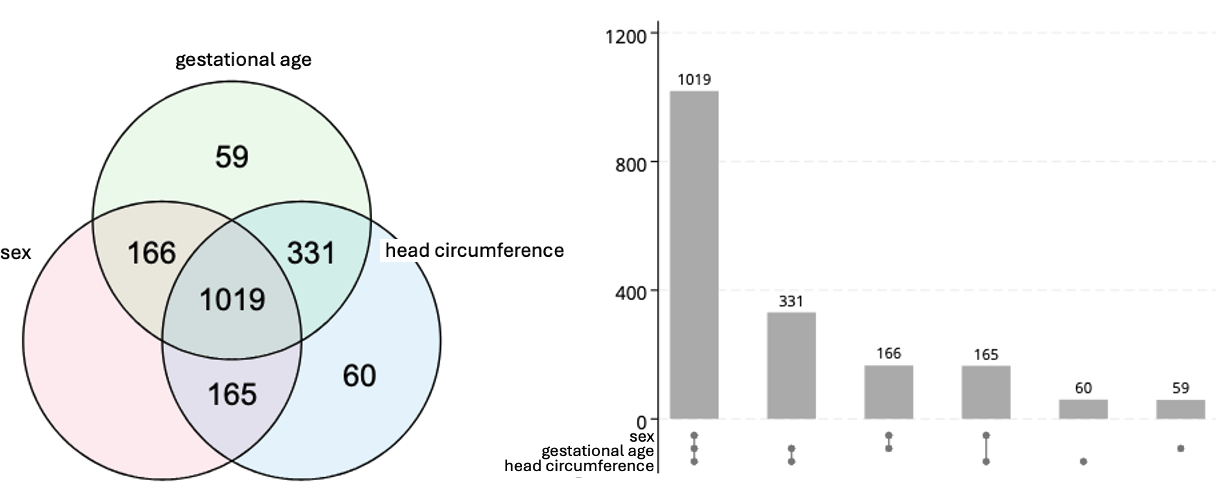}
    \caption{Venn diagram and UpSet plot indicating the number of observations in the dataset with observed data.   There are $n$=1,019 individuals for which there is complete data and $n$=781 individuals who have at least one value missing.}
    \label{fig:venn}
\end{figure}

There are 781 individuals who have at least one value missing; see Venn diagram in Figure \ref{fig:venn}.  Microcephaly status, $Y$, for a newborn can be calculated by first deriving their ``z-score'' based on their head circumference, gestational age and sex. (The  ``igb\_hcircm2zscore'' function in the growthstandards R package can do this easily using the \textsc{intergrowth} standards).  If a newborn's z-score is below $-2$, they are classified as having microcephaly.  As such, while the $f()$ function is known and can be easily applied to the source variables to obtain the binary derived variable, it is not easily written out.


We consider two models: (1) a simple ``Bernoulli model'', and (2) a ``Bernoulli skew-Normal mixture-Normal model'' (BsNmN model).  Both models were fit using Stan with $M=5,000$ draws (6,000 iterations and 1,000 burn-ins) \citep{carpenter2017stan}; Stan code is available in the Appendix.

\begin{figure}
    \centering
    \includegraphics[width=13cm]{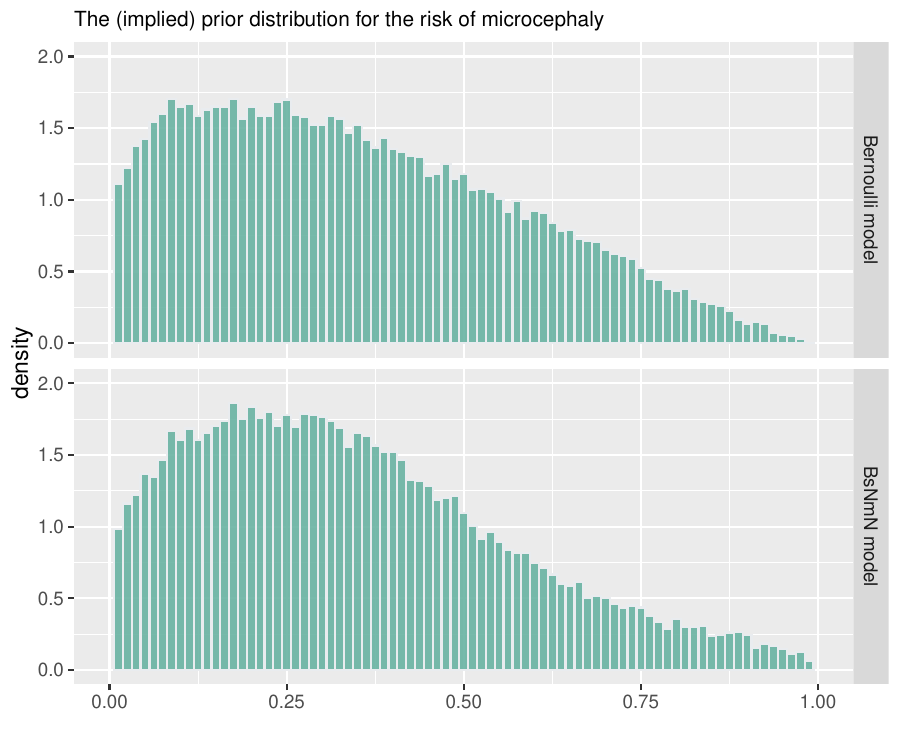}
    \caption{Lower panel shows histogram of a sample from the ``implied prior'' on $\textrm{E}(Y)$ used in the BsNmN model for the microcephaly example analysis. Upper panel shows histogram of a sample from the Beta(1.26, 2.32) prior placed on $\theta$ used in the Bernoulli model.  The numbers 1.26 and 2.32  were specifically chosen so that these two priors would be similar.}
    \label{fig:hist}
\end{figure}

The Bernoulli model involves considering the binary microcephaly status of each individual as the outcome of a simple Bernoulli model where  $\theta$ is the risk of microcephaly:
\begin{equation}
    Y \sim \textrm{Bernoulli}(\theta).
\end{equation}
We set a  $\textrm{Beta}(1.26, 2.32)$ prior for $\theta$ (see top panel of Figure \ref{fig:hist}), corresponding to an \textit{a priori} belief that  $\textrm{Pr}(\theta<31.5\%)=50\%$.   (This prior for was selected so that the two models have (implied) priors matching as closely as possible; see Figure \ref{fig:hist}).

We fit the Bernoulli model to the complete case data ($n=1,019$) and obtained an estimate of: $\hat{\theta} = 9.18\%$, 95\%CrI=  [7.60\%, 10.92\%]  (posterior median and equal-tailed 95\% credible interval).  We also fit the Bernoulli model to the entire dataset ($n$= 1,800) and followed Gelman's three-step procedure with SVL imputation.  Specifically, we used MICE specifying the ``logreg''  method for $Z_{1}$  and the ``normal'' method for $Z_{2}$ and $Z_{3}$.  For each of $K=50$ imputed datasets, we fit the Bernoulli model using MCMC with $M$ = 5,000 draws and after combining all 250,000 draws together, we obtained a posterior median estimate of: $\hat{\theta} = 12.70\%$, with equal-tailed 95\%CrI= [11.00\%, 14.51\%].

The BsNmN model involves a series of conditional distributions for each of the three source variables.  We assume that  an infant's sex ($Z_{1}$) is unlikely to impact their gestational age at birth ($Z_{2}$) (at least by any appreciable amount; see \cite{broere2016sex}).  Therefore, for sex, we define:

\begin{equation}
    Z_{1} \sim \textrm{Bernoulli}(0.5).
\end{equation}
\noindent  The distribution of gestational age ($Z_{2}$) is known to be skewed with a long left tail (due to preterm delivery), almost complete truncation on the right tail at 45 weeks (due to medically-induced labour at around 45 weeks) and almost complete truncation on the left tail at around 24 weeks (due to viability).  Many complex distributions have been suggested for modelling gestational age (e.g., \citet{rathjens2023bivariate} recommends the three-parameter Dagum distribution).  Following \citet{sauzet2015dichotomisation}, we choose to define a skew-Normal distribution centered at 39 weeks:
\begin{equation}
    (Z_{2}-39) \sim \textrm{skew-Normal}\left(\mu - \sqrt{\frac{2}{\pi}}\frac{\sigma
\omega}{\sqrt{1+\omega^2}}, \sigma, \omega\right),
\end{equation}
%
%

%
\noindent where $\mu$ is the mean of the distribution (known to be approximately 39 weeks, according to \citet{sauzet2015dichotomisation}), $\sigma$ is the scale, and $\omega$ is the slant.

Finally, head circumference ($Z_{3}$) is known to be approximately Gaussian and mean head circumference increases in an approximately quadratic way with gestational age (see Figure 2C in \citet{villar2014international}, and Figure \ref{fig:quad} in the Appendix). In a ZIKV infected population, infection is thought to cause a certain proportion of infants to have smaller heads.  As in \citet{kalmin2019misclassification}, we define a Normal mixture distribution for $Z_{3}$:
\begin{align}
    Z_{3}|Z_{1}=z_{1}, Z_{2}=z_{2}, X=1 \quad &\sim \nonumber \\
    \quad w\textrm{Normal}(\beta_{0} + \beta_{1}z_{1}& + \beta_{2}(z_{2}-39) + \beta_{3}(z_{2}-39)^{2}, \zeta_{1}) + \nonumber \\ (1-w)\textrm{Normal}(\kappa & + \beta_{1}z_{1} + \beta_{2}(z_{2}-39) + \beta_{3}(z_{2}-39)^{2}, \zeta_{2}),
\end{align}

\noindent where  $w$ is the proportion of ``non-affected'' individuals and $(1-w)$ is the proportion of ``affected'' individuals. The  $\beta_{0}$, $\beta_{1}$, $\beta_{2}$ and $\zeta_{1}$ parameters relate the distribution of head circumference to sex and gestational age for the ``non-affected'' population; and the $\kappa$ and $\zeta_{2}$ parameters correspond to the impact of ZIKV for the ``affected'' individuals in terms of the difference in mean head circumference and the scale.

The parameters in the BsNmN model are $\psi=(\mu, \sigma, \omega, \beta_{0}, \beta_{1}, \beta_{2},  \beta_{3}, \zeta_{1},\zeta_{2}, \kappa, w)$ and each requires a prior.  Based on the information available in \citet{villar2014international} and \citet{sauzet2015dichotomisation}, we can be reasonably certain of how gestational age and head circumference are distributed in the non-affected population and therefore set the following informative priors:
\begin{align}
&\mu \sim \textrm{Normal}(0, 0.1),\nonumber \\
&\beta_{0} \sim \textrm{Normal}(33.912, 0.1), \quad
&\beta_{1} \sim \textrm{Normal}(-0.450, 0.1), \nonumber \\
&\beta_{2} \sim \textrm{Normal}(0.399, 0.1),\quad
\textrm{and} 
&\beta_{3} \sim \textrm{Normal}(-0.016, 0.1). \nonumber   
\end{align}
For the scale and slant parameters, we set the following weakly informative priors:
\begin{align}
&\sigma  \sim \textrm{inv-Gamma}(2, 2), &\omega \sim \textrm{Normal}(0, 2), \nonumber \\
&\zeta_{1} \sim \textrm{inv-Gamma}(2, 2), \textrm{and} &\zeta_{2} \sim \textrm{inv-Gamma}(2, 2). \nonumber 
\end{align}
Finally, a truncated Normal prior for $\kappa$ was chosen in an effort to help address issues with identifiability that are common when fitting Bayesian mixture models \citep{Betancourt}, and a uniform prior was chosen for $w$:
\begin{align}
\kappa  &\sim \textrm{Normal}(-2, 2)_{[,-1]}, \nonumber 
\quad \quad w \sim \textrm{Uniform}(0, 1).
\end{align}

\begin{figure}
    \centering
    \includegraphics[width=14cm]{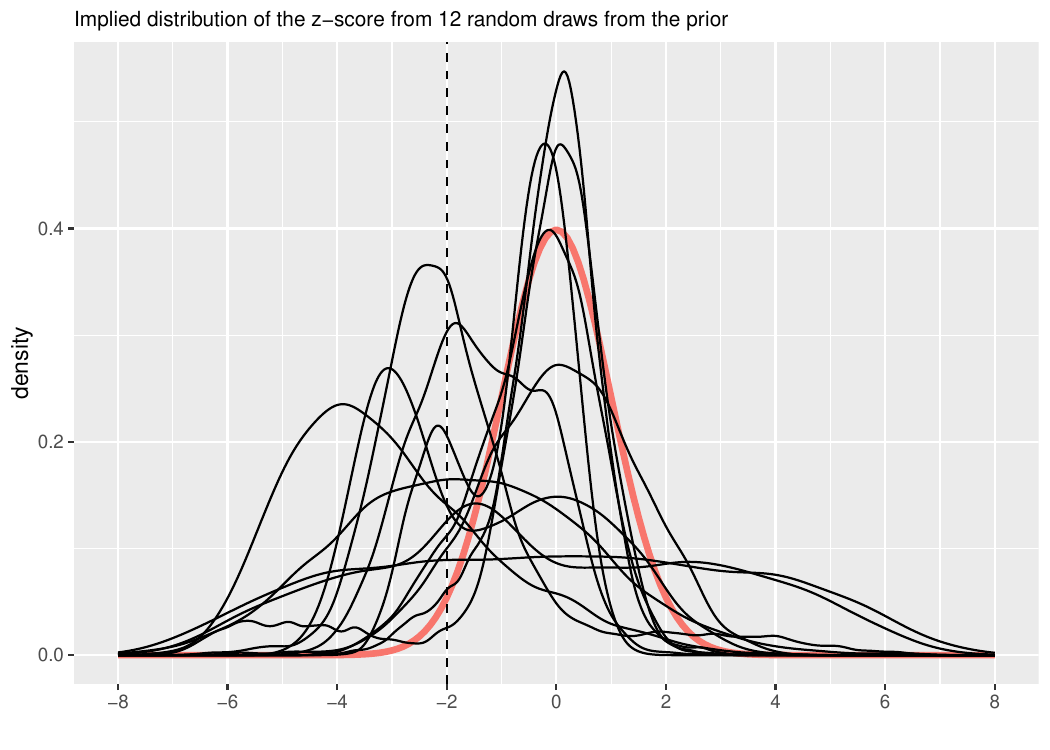}
    \caption{The implied distribution of the z-score from 12 random draws from the prior of the BsNmN model for the microcephaly example analysis. For reference, the red curve corresponds to the density of the standard normal.}
    \label{fig:priorZ}
\end{figure}

These priors correspond to an \textit{a priori} belief that $\textrm{Pr}(\textrm{E}(Y)<31.5\%)=0.5$.  The bottom panel of Figure \ref{fig:hist} plots the implied prior distribution for $\textrm{E}(Y)$ (i.e., for the risk of microcephaly) and Figure \ref{fig:priorZ} plots the implied distribution of the z-score from 12 random draws of the prior.

Fitting the BsNmN model to the complete case data ($n=1,019$) we obtain an estimate of: $\hat{\theta} = 9.18\%$, 95\%CrI=  [7.60\%,  10.92\%].  We also applied the proposed approach to fit the BsNmN model to the entire dataset ($n=1,800$) as detailed in Section 2 with $S$=5,000 and obtained an estimate of: $\hat{\theta} = 11.68\%$, 95\%CrI= [10.16\%, 13.36\%].   Figure \ref{fig:posterior} plots the implied distribution of the z-score from the prior and posterior.  Figures \ref{fig:tracecomplete} and \ref{fig:tracefull} in the Appendix show the trace plots of the MCMC.

\begin{figure}
    \centering
    \includegraphics[width=\linewidth]{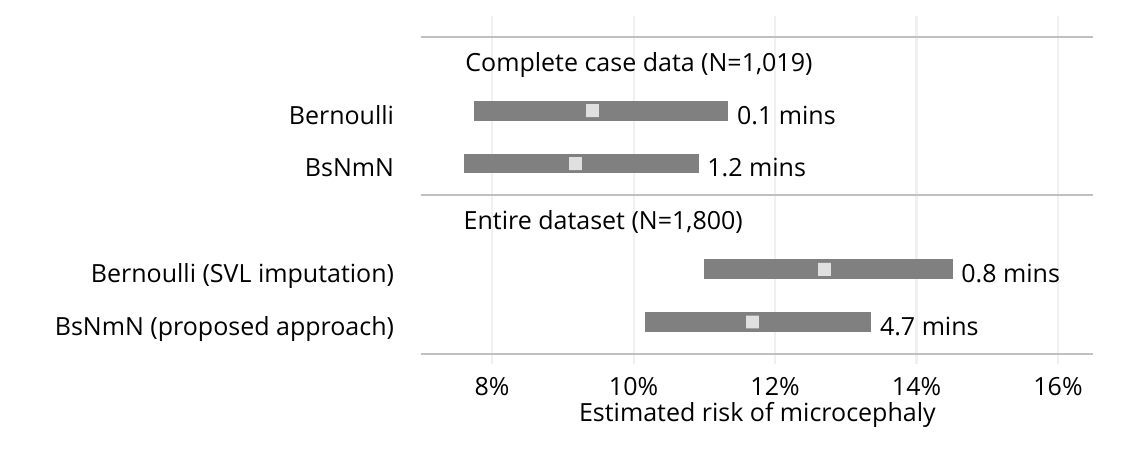}
    \caption{Results from the microcephaly analyses (posterior medians and equal-tailed 95\% credible intervals).}
    \label{fig:results}
\end{figure}

       

         


\begin{figure}
    \centering
    \includegraphics[width=15.5cm]{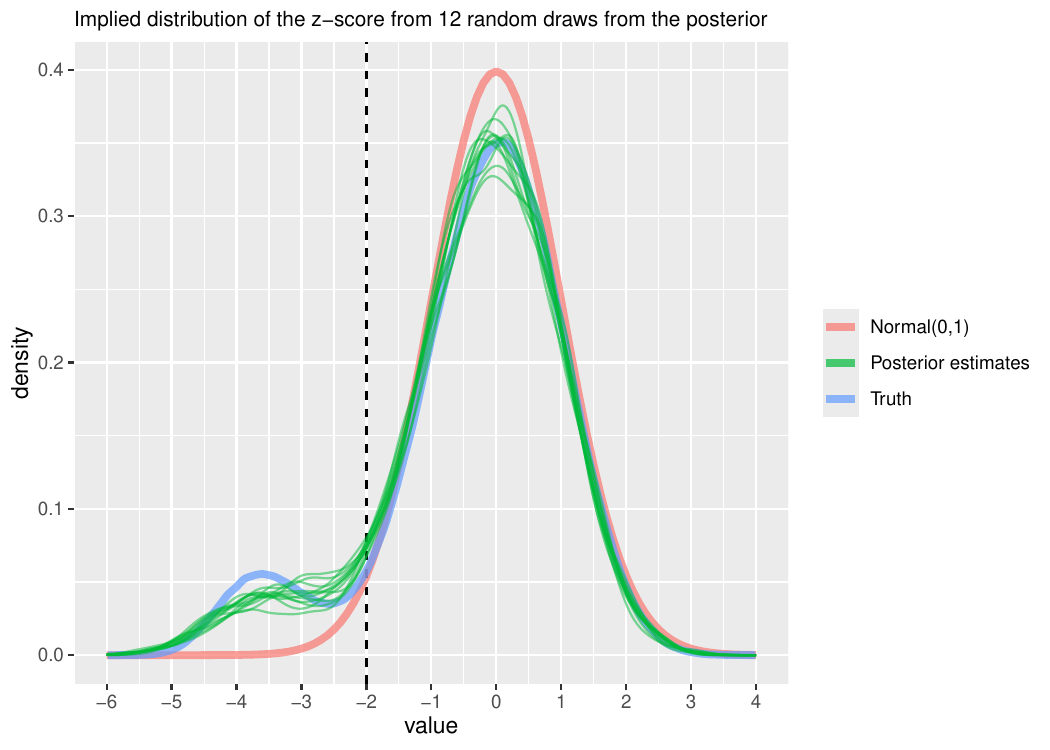}
    \caption{The implied distribution of the z-score from 12 random draws from the posterior of the BsNmN model.  Newborns with z-scores less than $-2$ (vertical dashed line) are classified as having microcephaly. For reference, the red curve corresponds to the density of the standard normal and the blue curve corresponds to the ``truth'' from which the data were simulated.}
    \label{fig:posterior}
\end{figure}

\section{Discussion}

If the amount of missingness is very small or negligible, applying a missing value procedure should not drastically change one's estimates, even if this procedure involves a misspecified model for the missing values.  Within Gelman's three step method, if the amount of missingness is negligible, an incorrect imputation model should have only a minimal negative effect.  In our proposed fully-integrated Bayesian approach, the model for the missing values and the outcome model are one and the same.  Therefore if this model is misspecified, the consequences might be more substantial even if there is negligible missingness.  This model might also be more complex than the one originally considered for analysis.  For instance, a researcher might be able to make relatively simple assumptions about the distribution of the derived outcome variable (e.g., be able to assume that microcephaly status is binomial), but have difficulty in making reasonable distributional assumptions for the source variables (e.g., what are  reasonable distributions to assume for gestational age and head circumference?).

The advantages of modeling the source variables directly include the ability to incorporate informed priors specific to the individual source variables and model missingness within a fully Bayesian model following our proposed method.  One could also incorporate adjustments for other sources of bias beyond missingness such as preferential sampling \citep{campbell2022bayesian} and measurement error.  This is an important future goal for the microcephaly model given the issues considered by \citet{harville2020measurement}.  

Finally, we note that our proposed approach allows one to conduct the analysis of data with missingness within an entirely Bayesian framework.  In some instances, one might wish to fit a Bayesian imputation model for the missing values and frequentist model for the main analysis (i.e., for the outcome model).  In such a case, Rubin's rules could be appropriately and easily applied.  Alternatively, if both the imputation model and the outcome model of interest must be frequentist, von Hippel \& Bartlett's rules \citep{von2021maximum} may be more appropriate.  To the best of our knowledge, if the imputation model is frequentist and the outcome model is Bayesian (an admittedly obscure scenario), there does not appear to be a good approach for obtaining estimates.

\bibliography{main}

@article{rubin1976inference,
  title={Inference and missing data},
  author={Rubin, Donald B},
  journal={Biometrika},
  volume={63},
  number={3},
  pages={581--592},
  year={1976},
  publisher={Oxford University Press}
}

@article{harville2020measurement,
  title={Measurement error, microcephaly prevalence and implications for {Z}ika: {A}n analysis of {U}ruguay perinatal data},
  author={Harville, Emily W and Buekens, Pierre M and Cafferata, Maria Luisa and Gilboa, Suzanne and Tomasso, Giselle and Tong, Van},
  journal={Archives of {D}isease in {C}hildhood},
  volume={105},
  number={5},
  pages={428--432},
  year={2020},
  publisher={BMJ Publishing Group Ltd}
}

@article{carpenter2017stan,
  title={Stan: A probabilistic programming language},
  author={Carpenter, Bob and Gelman, Andrew and Hoffman, Matthew D and Lee, Daniel and Goodrich, Ben and Betancourt, Michael and Brubaker, Marcus A and Guo, Jiqiang and Li, Peter and Riddell, Allen},
  journal={Journal of {S}tatistical {S}oftware},
  volume={76},
  year={2017},
  publisher={NIH Public Access}
}

@article{wilder2019understanding,
  title={Understanding the relation between Zika virus infection during pregnancy and adverse fetal, infant and child outcomes: a protocol for a systematic review and individual participant data meta-analysis of longitudinal studies of pregnant women and their infants and children},
  author={Wilder-Smith, Annelies and Wei, Yinghui and de Ara{\'u}jo, Thalia Velho Barreto and VanKerkhove, Maria and Martelli, Celina Maria Turchi and Turchi, Mar{\'\i}lia Dalva and Teixeira, Mauro and Tami, Adriana and Souza, Jo{\~a}o and Sousa, Patricia and others},
  journal={BMJ {O}pen},
  volume={9},
  number={6},
  pages={e026092},
  year={2019},
  publisher={British Medical Journal Publishing Group}
}

@article{kahan2023eliminating,
  title={Eliminating ambiguous treatment effects using estimands},
  author={Kahan, Brennan C and Cro, Suzie and Li, Fan and Harhay, Michael O},
  journal={American {J}ournal of {E}pidemiology},
  volume={192},
  number={6},
  pages={987--994},
  year={2023},
  publisher={Oxford University Press}
}

@article{von2021maximum,
  title={Maximum Likelihood Multiple Imputation},
  author={Von Hippel, Paul T and Bartlett, Jonathan W},
  journal={Statistical Science},
  volume={36},
  number={3},
  pages={400--420},
  year={2021},
  publisher={JSTOR}
}

@book{kuczmarski2000cdc,
  title={CDC growth charts: United States},
  author={Kuczmarski, Robert J},
  number={314},
  year={2000},
  publisher={US Department of Health and Human Services, Centers for Disease Control and~…}
}

@article{gustaf2015,
author = {Gustafson, Paul},
title = {Discussion of “On Bayesian Estimation of Marginal Structural Models”},
journal = {Biometrics},
volume = {71},
number = {2},
pages = {291-293},
doi = {https://doi.org/10.1111/biom.12271},
url = {https://onlinelibrary.wiley.com/doi/abs/10.1111/biom.12271},
eprint = {https://onlinelibrary.wiley.com/doi/pdf/10.1111/biom.12271},
year = {2015}
}

@article{tilling2016appropriate,
  title={Appropriate inclusion of interactions was needed to avoid bias in multiple imputation},
  author={Tilling, Kate and Williamson, Elizabeth J and Spratt, Michael and Sterne, Jonathan AC and Carpenter, James R},
  journal={Journal of clinical epidemiology},
  volume={80},
  pages={107--115},
  year={2016},
  publisher={Elsevier}
}

@article{keil2018bayesian,
  title={A {B}ayesian approach to the g-formula},
  author={Keil, Alexander P and Daza, Eric J and Engel, Stephanie M and Buckley, Jessie P and Edwards, Jessie K},
  journal={Statistical {M}ethods in {M}edical {R}esearch},
  volume={27},
  number={10},
  pages={3183--3204},
  year={2018},
  publisher={SAGE Publications Sage UK: London, England}
}

@article{vansteelandt2011invited,
  title={Invited commentary: G-computation--lost in translation?},
  author={Vansteelandt, Stijn and Keiding, Niels},
  journal={American journal of epidemiology},
  volume={173},
  number={7},
  pages={739--742},
  year={2011},
  publisher={Oxford University Press}
}

@article{zou2009assessment,
  title={Assessment of risks by predicting counterfactuals},
  author={Zou, GY},
  journal={Statistics in medicine},
  volume={28},
  number={30},
  pages={3761--3781},
  year={2009},
  publisher={Wiley Online Library}
}

@article{plummer2004jags,
  title={JAGS: Just another Gibbs sampler},
  author={Plummer, Martyn},
  year={2004}
}

@article{van2015package,
  title={Package ‘mice’},
  author={van Buuren, Stef and Groothuis-Oudshoorn, Karin and Robitzsch, Alexander and Vink, Gerko and Doove, Lisa and Jolani, Shahab and others},
  journal={Computer {S}oftware},
  year={2015},
  publisher={Citeseer}
}

@article{broere2016sex,
  title={Sex-specific differences in fetal and infant growth patterns: a prospective population-based cohort study},
  author={Broere-Brown, Zoe A and Baan, Esme and Schalekamp-Timmermans, Sarah and Verburg, Bero O and Jaddoe, Vincent WV and Steegers, Eric AP},
  journal={Biology of {S}ex {D}ifferences},
  volume={7},
  number={1},
  pages={1--9},
  year={2016},
  publisher={BioMed Central}
}

@article{rathjens2023bivariate,
  title={Bivariate Analysis of Birth Weight and Gestational Age by {B}ayesian Distributional Regression with Copulas},
  author={Rathjens, Jonathan and Kolbe, Arthur and H{\"o}lzer, J{\"u}rgen and Ickstadt, Katja and Klein, Nadja},
  journal={Statistics in {B}iosciences},
  pages={1--28},
  year={2023},
  publisher={Springer}
}

@article{zika2020zika,
  title={The {Z}ika virus individual participant data Consortium: {A} global initiative to estimate the effects of exposure to {Z}ika virus during pregnancy on adverse fetal, infant, and child health outcomes},
  author={{Zika Virus Individual Participant Data Consortium}},
  journal={Tropical {M}edicine and {I}nfectious {D}isease},
  volume={5},
  number={4},
  pages={152},
  year={2020},
  publisher={MDPI}
}

@article{ludtke2020regression,
  title={Regression models involving nonlinear effects with missing data: A sequential modeling approach using Bayesian estimation.},
  author={L{\"u}dtke, Oliver and Robitzsch, Alexander and West, Stephen G},
  journal={Psychological methods},
  volume={25},
  number={2},
  pages={157},
  year={2020},
  publisher={American Psychological Association}
}

@article{seaman2012multiple,
  title={Multiple imputation of missing covariates with non-linear effects and interactions: an evaluation of statistical methods},
  author={Seaman, Shaun R and Bartlett, Jonathan W and White, Ian R},
  journal={BMC medical research methodology},
  volume={12},
  pages={1--13},
  year={2012},
  publisher={Springer}
}

@article{white2011multiple,
  title={Multiple imputation using chained equations: issues and guidance for practice},
  author={White, Ian R and Royston, Patrick and Wood, Angela M},
  journal={Statistics in medicine},
  volume={30},
  number={4},
  pages={377--399},
  year={2011},
  publisher={Wiley Online Library}
}

@article{von2009impute,
  title={How to impute interactions, squares, and other transformed variables},
  author={Von Hippel, Paul T},
  journal={Sociological methodology},
  volume={39},
  number={1},
  pages={265--291},
  year={2009},
  publisher={Wiley Online Library}
}

@article{Erler2021,
  doi = {10.18637/jss.v100.i20},
  year = {2021},
  publisher = {Foundation for Open Access Statistic},
  volume = {100},
  number = {20},
  author = {Nicole S. Erler and Dimitris Rizopoulos and Emmanuel M. E. H. Lesaffre},
  title = {{JointAI}: Joint Analysis and Imputation of Incomplete Data in {R}},
  journal = {Journal of {S}tatistical {S}oftware}
}

@article{reiter2007multiple,
  title={The multiple adaptations of multiple imputation},
  author={Reiter, Jerome P and Raghunathan, Trivellore E},
  journal={Journal of the American Statistical Association},
  volume={102},
  number={480},
  pages={1462--1471},
  year={2007},
  publisher={Taylor \& Francis}
}

@article{pan2020passive,
  title={A passive and inclusive strategy to impute missing values of a composite categorical variable with an application to determine {H}{I}{V} transmission categories},
  author={Pan, Yi and He, Yulei and Song, Ruiguang and Wang, Guoshen and An, Qian},
  journal={Annals of {E}pidemiology},
  volume={51},
  pages={41--47},
  year={2020},
  publisher={Elsevier}
}

@book{gelman2004bayesian,
  title={Bayesian {D}ata {A}nalysis},
  author={Gelman, Andrew and Carlin, John B and Stern, Hal S and Dunson, David B and Vehtari, Aki and Rubin, Donald B},
  year={2004},
  publisher={CRC press}
}

@article{zhou2010note,
  title={A note on {B}ayesian inference after multiple imputation},
  author={Zhou, Xiang and Reiter, Jerome P},
  journal={The {A}merican {S}tatistician},
  volume={64},
  number={2},
  pages={159--163},
  year={2010},
  publisher={Taylor \& Francis}
}

@article{villar2014international,
  title={International standards for newborn weight, length, and head circumference by gestational age and sex: the Newborn Cross-Sectional Study of the {I}{N}{T}{E}{R}{G}{R}{O}{W}{T}{H}-21st {P}roject},
  author={Villar, Jos{\'e} and Ismail, Leila Cheikh and Victora, Cesar G and Ohuma, Eric O and Bertino, Enrico and Altman, Doug G and Lambert, Ann and Papageorghiou, Aris T and Carvalho, Maria and Jaffer, Yasmin A and others},
  journal={The {L}ancet},
  volume={384},
  number={9946},
  pages={857--868},
  year={2014},
  publisher={Elsevier}
}

@article{morris2014multiple,
  title={Multiple imputation for an incomplete covariate that is a ratio},
  author={Morris, Tim P and White, Ian R and Royston, Patrick and Seaman, Shaun R and Wood, Angela M},
  journal={Statistics in medicine},
  volume={33},
  number={1},
  pages={88--104},
  year={2014},
  publisher={Wiley Online Library}
}

@article{gottschall2012comparison,
  title={A comparison of item-level and scale-level multiple imputation for questionnaire batteries},
  author={Gottschall, Amanda C and West, Stephen G and Enders, Craig K},
  journal={Multivariate {B}ehavioral {R}esearch},
  volume={47},
  number={1},
  pages={1--25},
  year={2012},
  publisher={Taylor \& Francis}
}

@article{campbell2022bayesian,
  title={Bayesian adjustment for preferential testing in estimating infection fatality rates, as motivated by the COVID-19 pandemic},
  author={Campbell, Harlan and de Valpine, Perry and Maxwell, Lauren and de Jong, Valentijn MT and Debray, Thomas PA and Jaenisch, Thomas and Gustafson, Paul},
  journal={The {A}nnals of {A}pplied {S}tatistics},
  volume={16},
  number={1},
  pages={436--459},
  year={2022},
  publisher={Institute of Mathematical Statistics}
}

@article{Betancourt,
    title = {Identifying Bayesian mixture models},
    author = {Betancourt, M.},
year= {2017},
journal={Stan Case Studies, \url{mc-stan.org/
users/documentation/case-studies/identifying_mixture_models.html}}
}

@article{kalmin2019misclassification,
  title={Misclassification in defining and diagnosing microcephaly},
  author={Kalmin, Mariah M and Gower, Emily W and Stringer, Elizabeth M and Bowman, Natalie M and Rogawski McQuade, Elizabeth T and Westreich, Daniel},
  journal={Paediatric and {P}erinatal {E}pidemiology},
  volume={33},
  number={4},
  pages={286--290},
  year={2019},
  publisher={Wiley Online Library}
}

@article{sauzet2015dichotomisation,
  title={Dichotomisation using a distributional approach when the outcome is skewed},
  author={Sauzet, Odile and Ofuya, Mercy and Peacock, Janet L},
  journal={BMC {M}edical {R}esearch {M}ethodology},
  volume={15},
  pages={1--11},
  year={2015},
  publisher={Springer}
}

@book{van2018flexible,
  title={Flexible imputation of missing data},
  author={Van Buuren, Stef},
  year={2018},
  publisher={CRC press}
}

@article{bartlett2015multiple,
  title={Multiple imputation of covariates by fully conditional specification: accommodating the substantive model},
  author={Bartlett, Jonathan W and Seaman, Shaun R and White, Ian R and Carpenter, James R and {Alzheimer's Disease Neuroimaging Initiative*}},
  journal={Statistical {M}ethods in {M}edical {R}esearch},
  volume={24},
  number={4},
  pages={462--487},
  year={2015},
  publisher={Sage Publications Sage UK: London, England}
}

@article{Pham2021,
  author = {T. My~Pham and I. R. White and B. C. Kahan and T. P. Morris and S. J. Stanworth and G. Forbes},
  title = {A comparison of methods for analyzing a binary composite endpoint with partially observed components in randomized controlled trials},
  journal = {Statistics in Medicine},
  doi = {10.1002/sim.9203},
  year = {2021},
  publisher = {Wiley},
  volume = {40},
  number = {29},
  pages = {6634--6650}
}

\section{Appendix}

\subsection*{Simulation study}

Let us now review the results of a small simulation study to confirm that our intuitions regarding each of the possible strategies in the illustrative example. We simulated 500 datasets of size $n=1,000$ and applied each of the six methods (details are in the Appendix).  The aim was to demonstrate a proof-of-concept for our proposed method. 

Five hundred complete-data observations were generated for group A and another five hundred complete-data observations were generated for group B as follows.  For $i$ in 1,...,$n$

\begin{align*}
\left(\begin{array}{c}Z_1 \\ Z_2 \end{array}\right) & \sim 
\textrm{Normal}\begin{pmatrix}  
\left(\begin{array}{c}1 \\ 2 \end{array}\right), \left(\begin{array}{cc}1 & 0.25 \\ 0.25 & 1 \end{array}\right) \end{pmatrix},& \text{if $i$ is in group $A$} \\ 
\left(\begin{array}{c}Z_1 \\ Z_2 \end{array}\right) & \sim 
\textrm{Normal}\begin{pmatrix}  \left(\begin{array}{c}1.5 \\ 1 \end{array}\right), \left(\begin{array}{cc}1 & 0.25 \\ 0.25 & 1 \end{array}\right) \end{pmatrix},& \text{if $i$ is in group $B$}.  \\
\end{align*}

\noindent As in the illustrative example, the derived outcome variable is $Y = Z_{1} + Z_{2}$ and the interest is in estimating $\theta$, the difference between the expectation of $Y$ in group~$A$ and in group~$B$.  The true value of $\theta=-0.5$. 

We analysed the data prior to introduction of any missingness (i.e., from the ``oracle'') with the univariate normal model (see \ref{eq:univariate}).  Missingness was then imposed under a missing-at-random mechanism as follows. All data remained fully observed for group~$B$. For group~$A$, a randomly selected 250 observations had non-zero probability of $z_1$ being missing, while the other 250 had non-zero probability of $z_2$ being missing. The following models determined the probability of missingness in these two subsets of the group~$A$ data:
\begin{align*}
\text{logit}[P(z_1 \textrm{ is missing})] &= 10 + 10z_2,\\
\text{and~~logit}[P(z_2 \textrm{ is missing})] &= 4 - 5z_1.
\end{align*}

For analysis of the incomplete data, each of the MICE methods was used to produce 50 imputed datasets, each Bayesian model was fit using MCMC with $M=5,000$ draws, and the proposed method was implemented with $S=50,000$ Monte Carlo samples for the integration.

\begin{figure}
    \centering
    \includegraphics[width=14cm]{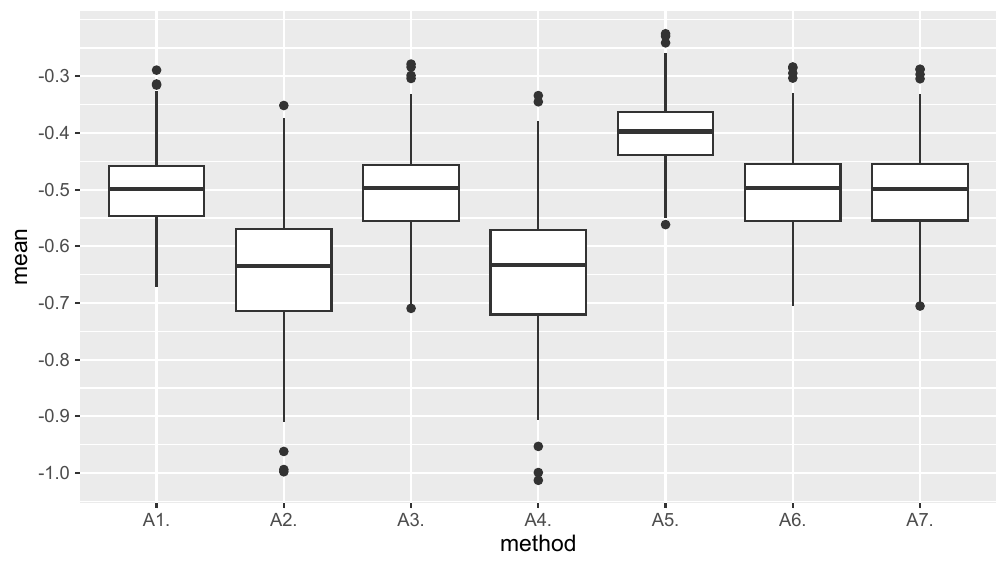}
    \caption{Simulation results. We simulated 500 datasets of size $n=1,000$ with true $\theta=-0.5$ and applied each of the six methods (A2--A7) recording the posterior mean estimate.   We also analyzed data before introducing missingness (\textit{i.e.}, from the oracle) and fit the univariate normal model to these data. For reference:  A1. = Oracle with univariate Normal model; A2. = Complete case analysis with univariate Normal model; A3. = Gelman's 3 step with SVL imputation and univariate Normal model; A4. = Gelman's 3 step with DVL imputation with univariate Normal model; A5. = Gelman's 3 step with JAV imputation with univariate Normal model; A6. = Bivariate Normal model (with math); and A7. = Bivariate Normal model (with MC approx.).}
    \label{fig:sim1}
\end{figure}

Figure \ref{fig:sim1} plots the simulation study results.  In summary, the we see that:
\begin{itemize}
    \item The complete-case analysis (A2) and the three-step approach with DVL imputation (A4) are both biased as expected, since the data generating mechanism is not MCAR.
    \item The three-step approach with SVL imputation (A3) is unbiased.
    \item Both approaches that fit the bivariate Normal model (A6 and A7) are unbiased and just as efficient as the three-step approach with SVL imputation (A3).
    \item Finally, the three-step approach with JAV (A5) imputation appears to be biased. This is surprising, since in this case JAV imputation should be identical to SVL imputation. A brief explanation of this claim is given in the Appendix. In short, this is not a property of JAV imputation but of the R package \{mice\} used to implement it.
\end{itemize}

In the simulation study, the three-step approach with JAV (A5) imputation appears to be biased. This is surprising, since in this case the JAV imputation should be identical to the SVL imputation. 
 
Recall that JAV imputation used a fully-conditional specification procedure -- specifically the chained equations flavour -- rather than a joint model. Denote a random draw with $^*$ and the $c$th cycle using the superscript $^{(c)}$. The procedure (typically) initiates in a monotone fashion, visiting variables from least to most missing data, omitting incomplete variables that have not yet been imputed at the zeroth cycle. By construction, $Y$ always has more missing data than $Z_1$ or $Z_2$ and, in a given repetition, either $Z_1$ or $Z_1$ may have the least. For illustration, suppose that $Z_1$ has the least missing data. The procedure then draws
    \begin{align*}
        Z_1^{*(0)} &| \text{Group},\\
        Z_2^{*(0)} &| Z_1^{*(0)},Z_1^{(obs)},\text{Group}. \\
    \end{align*}
    So far, this is identical to SVL imputation. Next, the procedure aims to draw
    \begin{align*}
            Y^{*(0)} &| Z_1^{*(0)},Z_1^{(obs)},Z_2^{*(0)},Z_2^{(obs)},\text{Group}.
    \end{align*}
To do so, it starts by fitting a linear regression model for $Y$ conditioning on $Z_1^{(obs)},Z_2^{(obs)},\text{Group}$ (note that $Z_1^{*(1)},Z_2^{*(1)}$ are not conditioned-on because chained equations fits the imputation model for a given variable among those with that variable observed, which makes it more efficient than a Gibbs sampler). Now that $Y$ conditions on the variables that perfectly predict its value, the linear regression model will correctly estimate the coefficients, without uncertainty and without residual error. Even if it draws $Y^{*(0)}$ then $Z_1^{*(1)},Z_1^{*(1)}$ and so on, the final imputations have already been completely determined by $Z_1^{*(0)}$ and $Z_2^{*(0)}$. Thus, the procedure should be identical to SVL imputation. Where does the apparent bias come from? We believe it is due to default choices made by the R package \texttt{mice} (possibly to handle perfect prediction, or drop variables; we are unsure). Indeed, to check this, we re-ran the simulation study using Stata's \texttt{mi impute chained} to perform the imputation and, as expected, found that the JAV procedure described above was unbiased as expected.

\subsection*{Math derivation for Dutch boys example}

Note that:
\begin{equation}
Y = \textrm{log}\Big(\frac{Z_{2}}{({Z_{1}/100})^{2}}\Big)
    = \textrm{log}(Z_{2}) - 2\textrm{log}(Z_{1}) + 2\textrm{log}(100)
\end{equation}
so that, for the bivariate model we have:
\begin{equation}
Y \sim \textrm{Normal}(\gamma_{0} + \gamma_{1}r + \gamma_{2}a + \gamma_{3}ra  -2(\alpha_{0} + \alpha_{1}r + \alpha_{2}a + \alpha_{3}ra) + 2\textrm{log}(100), \tau_{Z2}^2 + 4\tau_{Z1}^2 + 2\tau_{Z2}\tau_{Z1}\rho) 
\end{equation}
and:
\begin{align}
\theta &= \int_{a}\textrm{E}(Y|R=1, A=a)f_{A}(a)\textrm{da} - \int_{a}\textrm{E}(Y|R=0, A=a)f_{A}(a)\textrm{da} \\ \nonumber
&= \int_{a} \gamma_{0} + \gamma_{1}1 + \gamma_{2}a + \gamma_{3}1a + 2\textrm{log}(100) -2(\alpha_{0} + \alpha_{1}1 + \alpha_{2}a + \alpha_{3}1a) - \\ \nonumber
& \quad (\gamma_{0}  + \gamma_{2}a + 2\textrm{log}(100) -2(\alpha_{0} + \alpha_{2}a)) \textrm{da}\\
&= \int_{a}{(\gamma_{1} + \gamma_{3}a -2\alpha_{1} - 2\alpha_{3}a)f_{A}(a)}\textrm{da}\\
&=\gamma_{1} -2\alpha_{1} + (\gamma_{3} - 2\alpha_{3})\textrm{E}(A)
\end{align}

\subsection*{Code for Dutch boys example}
\begin{footnotesize}
\begin{verbatim}
rm(list = ls())
detach_package <- function(pkg, character.only = FALSE)
{
  if(!character.only)
  {
    pkg <- deparse(substitute(pkg))
  }
  search_item <- paste("package", pkg, sep = ":")
  while(search_item %in% search())
  {
    detach(search_item, unload = TRUE, character.only = TRUE)
  }
}
detach_package("mice")
library(mice)
library(rjags)
library(mvtnorm)
summary(boys$age)
dim(boys)
boys<-boys[boys[,"age"]>=1 & boys[,"age"]<=18,]

boys[1:20,]
dim(boys) 

summary(boys$age)
boys$logbmi <- log(boys$bmi)
boys$loghgt <- log(boys$hgt)
boys$logwgt <- log(boys$wgt)
boys$city <- as.numeric(boys$reg=="city")
boys<-(boys[,c("hgt", "wgt", "logbmi", "loghgt", "logwgt", "city", "age")])

boxplot(boys$logbmi~boys$reg)

library(ggvenn)
ggvenn(as.data.frame(is.na(boys[,c("logbmi","loghgt","logwgt","city")])), 
fill_color=c("#0073C2FF", "#EFC000FF", "#868686FF", "#CD534CFF"))

# Frequentist analysis:
plot(logbmi~age,data=boys[boys[,"age"]>1,], 
    col=boys[boys[,"age"]>1,"city"]+1, pch=20)
mod1 <- lm(logbmi~city*age + I(age^2) ,
    data=boys[boys[,"age"]>1,])
freq_complete <- round(c(coef(mod1)["city"], confint(mod1)["city",]),3)
freq_complete
summary(mod1)


##############################
nMCMC <- 2000
thedata <- boys[,c("logbmi", "hgt", "wgt", "city", "age")]
dim(thedata)




# Univariate analysis:
# model 1: a model with y is input
jags_univ <- "model {
# Priors
beta0 ~ dnorm(0,1)
beta1 ~ dnorm(0,1)
beta2 ~ dnorm(0,1)
beta3 ~ dnorm(0,1)
beta4 ~ dnorm(0,1)
sigma ~ dexp(1)

# Model	
for(i in 1:N){
	y[i]     ~ dnorm(beta0 + beta1*x1[i] + 
                        beta2*x2[i]+ beta3*x1[i]*x2[i] + beta4*(x2[i]^2),
	 1/(sigma_squared))}

# Output
sigma_squared <- sigma^2
}"




# Bivariate analysis:
# model 2: a model with z1 and z2 as input
jags_biv<- "model {
# Priors
alpha0 ~ dnorm(0,1)
alpha1 ~ dnorm(0,1)
alpha2 ~ dnorm(0,1)
alpha3 ~ dnorm(0,1)
alpha4 ~ dnorm(0,1)

gamma0 ~ dnorm(0,1)
gamma1 ~ dnorm(0,1)
gamma2 ~ dnorm(0,1)
gamma3 ~ dnorm(0,1)
gamma4 ~ dnorm(0,1)

# Constructing the covariance matrix and the corresponding precision matrix.
    prec[1:2,1:2] <- inverse(cov[,])
    cov[1,1] <- sigma[1] * sigma[1]
    cov[1,2] <- sigma[1] * sigma[2] * rho
    cov[2,1] <- sigma[1] * sigma[2] * rho
    cov[2,2] <- sigma[2] * sigma[2]
    
# Flat priors on all parameters.
    sigma[1] ~ dexp(1) 
    sigma[2] ~ dexp(1) 
    rho ~ dunif(-1, 1)
    pi  ~ dunif(0, 1)
# Model	
for(i in 1:N){
  x1[i]  ~ dbin(pi,1);
  mu[i,1] = alpha0 + alpha1*x1[i] + 
  				alpha2*x2[i] + alpha3*x1[i]*x2[i] + alpha4*(x2[i]^2) ;
  mu[i,2] = gamma0 + gamma1*x1[i] + 
  				gamma2*x2[i] + gamma3*x1[i]*x2[i] + gamma4*(x2[i]^2);
	#z[i,1:2] ~ dmnorm(mu[i,1:2], prec[1:2,1:2]);
	
	
	z[i,1] ~ dmnorm(mu[i,1], pow(sigma[1],-2) );
	z[i,2] ~ dmnorm(mu[i,2] + (sigma[2]/sigma[1])*rho*(z[i,1]-mu[i,1]), 
	                1/((1-(rho^2))*(sigma[2]^2)) );
	
}

}"


############################################################
############################################################
### ### With univariate ### ###
start_time <- Sys.time()
# model with y as input
thedata_cc<-na.omit(thedata)
jags.m <- jags.model(textConnection(jags_univ), 
                     data = list(y = thedata_cc[,"logbmi"], 
                                 x1 = thedata_cc[,"city"],
                                 x2 = thedata_cc[,"age"],
                                 N = dim(thedata_cc)[1]))

# this is our estimate
mAsamples <-(coda.samples(jags.m, c("beta1", "beta3"),
    n.iter = nMCMC, n.burnin=1000))
theta_samples <- apply(mAsamples[[1]],1, function(q) {
q["beta1"] + q["beta3"]*mean(thedata[,"age"])})  

univariate_complete_est <- round(quantile(
        theta_samples, c(0.5,0.025,0.975)),3)
end_time <- Sys.time()
mAtime <- end_time-start_time
univariate_complete_time <-mAtime
univariate_complete_est
univariate_complete_time
############################################################
############################################################

####################
create_theta_gcomp <- function(q){

#	N <- dim(thedata)[1]
	N <- 1
	x1 = rep(0,S*N)
	x2 = sample(thedata[,"age"], S*N, replace=TRUE)
	
		mu1 = q["alpha0"] + q["alpha1"]*x1 + 
                q["alpha2"]*x2 + q["alpha3"]*x1*x2 + q["alpha4"]*(x2^2);
		mu2 = q["gamma0"] + q["gamma1"]*x1 + 
                q["gamma2"]*x2 + q["gamma3"]*x1*x2 + q["gamma4"]*(x2^2);

		Sigma = matrix(c(q["sigma[1]"]^2, 
		q["sigma[1]"]*q["sigma[2]"]*q["rho"],
		q["sigma[1]"]*q["sigma[2]"]*q["rho"],
		q["sigma[2]"]^2),2,2) 
	
	
logBMIstar0 <- apply(cbind(1:(S*N)),1, function(i){
			zstar_i <- rmvnorm(1,cbind(mu1[i], mu2[i]), sigma= Sigma)
			logBMIstar <- (zstar_i[2]) - log((exp(zstar_i[1])/100)^2)
			return(logBMIstar)})

# Y* ~ Y|X1=0,X2=samplex2
			
	x1 = rep(1,S*N)
	
		mu1 = q["alpha0"] + q["alpha1"]*x1 + 
                q["alpha2"]*x2 + q["alpha3"]*x1*x2 + q["alpha4"]*(x2^2);
		mu2 = q["gamma0"] + q["gamma1"]*x1 + 
                q["gamma2"]*x2 + q["gamma3"]*x1*x2 + q["gamma4"]*(x2^2);

	
logBMIstar1 <- apply(cbind(1:(S*N)),1, function(i){
			zstar_i <- rmvnorm(1,cbind(mu1[i], mu2[i]), sigma= Sigma)
			logBMIstar <- (zstar_i[2]) - log((exp(zstar_i[1])/100)^2)
			return(logBMIstar)})	
# Y* ~ Y|X1=1,X2=samplex2			
			
	thetasample <- mean(logBMIstar1)  -	mean(logBMIstar0)	
	
	return(thetasample)}
########################################


############################################################
### ### With bivariate and math and proposed ### ###
start_time <- Sys.time()
# model with z as input
jags.m <- jags.model(textConnection(jags_biv), 
                     data = list(z = cbind(log(thedata[,"hgt"]),
                                           log(thedata[,"wgt"])), 
                                 x1 = thedata[,"city"],
                                 x2 = thedata[,"age"],
                                 N = dim(thedata)[1]))

mAsamples <- coda.samples(jags.m, 
c("alpha0", "alpha1", "alpha2", "alpha3", "alpha4",
"gamma0", "gamma1", "gamma2", "gamma3", "gamma4", "sigma", "rho"),
n.iter = nMCMC, n.burnin=1000)
dim(mAsamples[[1]])[1]
end_time <- Sys.time()
biv_time <- end_time-start_time

#library(devtools)
#install_github("psolymos/pbapply")
library(pbapply)

# with math:
start_time <- Sys.time()
theta_samplesA <- apply(mAsamples[[1]],1, function(q) {
q["gamma1"] - 2*q["alpha1"] +
    (q["gamma3"]-2*q["alpha3"])*mean(thedata[,"age"])})  
math_est <- round(quantile(theta_samplesA, c(0.5,0.025,0.975)),3)
end_time <- Sys.time()
math_time <- (end_time-start_time) + biv_time 


# with gcomp:
start_time <- Sys.time()
x1 = thedata[,"city"]; x2 = thedata[,"age"]; S <- 2000
theta_samples <- pbapply(mAsamples[[1]], 1, create_theta_gcomp)
gcomp_est <- round(quantile(theta_samples, c(0.5,0.025,0.975)),3)
end_time <- Sys.time()
gcomp_time <- (end_time-start_time) + biv_time 





#########################
# Passive imputation strategy:



boys[,c("agecity")] <-(boys[,c("city")]*boys[,c("age")])
boys[,c("loghgt")] <- log(boys[,c("hgt")])
boys[,c("logwgt")] <- log(boys[,c("wgt")])
boys[,c("age_squared")] <- (boys[,c("age")])^2

thedata <- (boys[,c("logbmi", "loghgt", 
    "logwgt", "city", "age", "age_squared")])
dim(thedata)
start_time <- Sys.time()
dat <- thedata
init = mice(dat, maxit=0) 
head(dat)
meth = init$method
predM = init$predictorMatrix
meth
meth[c("loghgt")]="norm" 
meth[c("logwgt")]="norm" 
meth[c("age")]="norm" 
meth[c("city")]="pmm" 
meth[c("logbmi")]="~I(log( exp(logwgt)  /(( exp(loghgt)  /100)^2)))"
meth[c("age_squared")]="~I(age^2)"
meth[c("agecity")]=="~I(age*city)"
meth
n_imputations <- 50
imputed = mice(dat, method=meth, predictorMatrix=predM, m=n_imputations)
imputed_dat <- list()
for(j in 1:n_imputations){imputed_dat[[j]] <- complete(imputed,action=j)}

coda_samples <- NULL
for(j in 1:n_imputations){
  
  jags.m1 <- jags.model(textConnection(jags_univ), 
                        data = list(y = imputed_dat[[j]][,"logbmi"], 
                                    x1 = imputed_dat[[j]][,"city"],
                                    x2 = imputed_dat[[j]][,"age"],
                                    N = dim(imputed_dat[[j]])[1]))
                                                                        
  
  
  
  mAsamples <-(coda.samples(jags.m1, c("beta1", "beta3") , 
        n.iter = nMCMC, n.burnin=1000))
theta_samples <- apply(mAsamples[[1]],1, function(q) {
        q["beta1"] + q["beta3"]*mean(imputed_dat[[j]][,"age"])})  

  
  coda_samples <- c(coda_samples, theta_samples)
}
passive_samples <- coda_samples
c(mean(passive_samples), sd(passive_samples))
passive_est <- round(quantile(passive_samples, c(0.500,0.025,0.975)),3)
end_time <- Sys.time()
passive_time <- end_time-start_time


univariate_complete_est
as.numeric(univariate_complete_time[1])/60

passive_est
as.numeric(passive_time[1])/60

math_est
as.numeric(math_time[1])/60

gcomp_est
as.numeric(gcomp_time[1])/60



\end{verbatim}
\end{footnotesize}

\subsection{Stan code for ZIKV example}

\begin{footnotesize}
\begin{verbatim}

Bernoulli_model <- 
  "data {
  int<lower=0> N;     // Number of observations
  int<lower=0, upper=1> y[N];   // Binary outcome data
  real<lower=0> a;               // Beta distribution shape parameter
  real<lower=0> b;               // Beta distribution shape parameter
}

parameters {
  real<lower=0, upper=1> theta;  // Probability of success
}

model {
  // Likelihood
  for (i in 1:N) {
    y[i] ~ bernoulli(theta);
  }
  
  // Priors
  theta ~ beta(a, b);
}"



BsNmN_model <- "data {
  int<lower=0> N;          
  int<lower=0, upper=N> N_z1obs;
  int<lower=0, upper=N> N_z2obs;
  int<lower=0, upper=N> N_z3obs;
  int<lower=0, upper=N> N_z1mis;
  int<lower=0, upper=N> N_z2mis;
  int<lower=0, upper=N> N_z3mis;
  vector<lower=0, upper=1>[N_z1obs] z1_obs;  // sex variable
  vector[N_z2obs] z2_obs;             // gestational age variable
  vector[N_z3obs]  z3_obs;             // head circumf. variable
  int<lower=0, upper=1> z1mis_ind[N];
  int<lower=0, upper=N> ii_z1_obs[N_z1obs];
  int<lower=0, upper=N> ii_z2_obs[N_z2obs];
  int<lower=0, upper=N> ii_z3_obs[N_z3obs];
  int<lower=0, upper=N> ii_z1_mis[N_z1mis];
  int<lower=0, upper=N> ii_z2_mis[N_z2mis];
  int<lower=0, upper=N> ii_z3_mis[N_z3mis];
}

parameters {
  vector<lower=0, upper=1>[N_z1mis] z1_mis; 
  vector[N_z2mis]  z2_mis;
  vector[N_z3mis]  z3_mis;
  real<upper=-1> kappa;
  real beta01;
  vector<lower=0>[2] zeta;    
  real beta1;
  real beta2;
  real beta3;
  simplex[2] mixweight;
  
  real mu; // mean of X
  real<lower=0> sigma; // SD of X
  real omega; // shape of X
}

transformed parameters {
  vector[N] z1;  // sex variable
  vector[N] z2;  // gestational age variable
  vector[N] z3;  // head cir variable  
  z1[ii_z1_obs] = z1_obs;
  z1[ii_z1_mis] = z1_mis;
  z2[ii_z2_obs] = z2_obs;
  z2[ii_z2_mis] = z2_mis;
  z3[ii_z3_obs] = z3_obs;
  z3[ii_z3_mis] = z3_mis;  
  vector[2] beta0;
  real loc_x; // location of X
  real gm_x; // intermediate calculation for location and scale

  gm_x = sqrt(2/pi())*omega/sqrt(1+omega^2);
  loc_x = mu - sigma*gm_x/sqrt(1-gm_x^2);
  beta0[1] = beta01;
  beta0[2] = kappa;
}

model {
  // Prior distributions for GA
  mu ~ normal(0, 0.1);
  sigma ~ inv_gamma(2, 2); 
  omega ~ normal(0, 2);
   
  // Prior distributions for HC:  
  beta0[1] ~ normal(0, 0.1);
  beta0[2] ~ normal(-2, 2)T[,-1];
  zeta[1] ~ inv_gamma(2, 2); 
  zeta[2] ~ inv_gamma(2, 2); 
  beta1 ~ normal(-0.450, 0.1);
  beta2 ~ normal(0.399, 0.1);    
  beta3 ~ normal(-0.016, 0.1); 
      
  // Likelihood
 z1_mis ~ uniform(0,1);

 z2_obs ~ skew_normal(loc_x, sigma, omega);
 z2_mis ~ skew_normal(loc_x, sigma, omega);
 
 vector[2] log_mixweight = log(mixweight);
 for (n in 1:N) {
 	vector[2] lps = log_mixweight;
// for unknown sex
   if (z1mis_ind[n]==1 ) {
		    for (k in 1:2) {
 				lps[k] += log_mix( z1[n],
                    normal_lpdf( z3[n] | 33.912 + beta0[k] + beta1 + beta2*z2[n] + 
                    beta3*pow(z2[n],2), zeta[k]),
                    normal_lpdf( z3[n] | 33.912 + beta0[k]  + beta2*z2[n]+ 
                    beta3*pow(z2[n],2), zeta[k]));
    		}	    		
  	}
// for known sex  	
	else {
			for (k in 1:2) {
		      lps[k] += normal_lpdf(z3[n] | 33.912 + beta0[k] + 
		      	beta1*z1[n] + beta2*z2[n]+ 
                    beta3*pow(z2[n],2), zeta[k]);
    		}  	 	
  	}
  	target += log_sum_exp(lps);
  }
}
"

\end{verbatim}

\begin{figure}
    \centering
    \includegraphics[width=14cm]{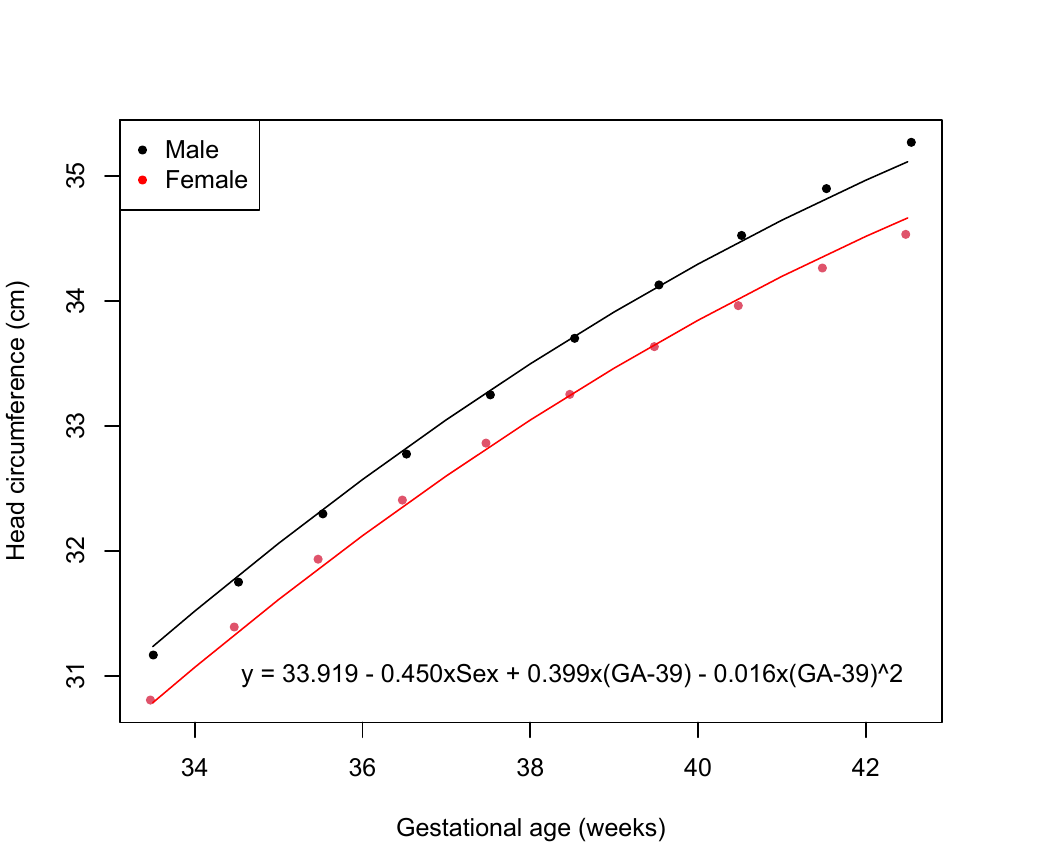}
    \caption{The quadratic regression model fit to digitized data from \citet{villar2014international}'s Figure 2C.}
    \label{fig:quad}
\end{figure}

\begin{figure}
    \centering
    \includegraphics[width=14cm]{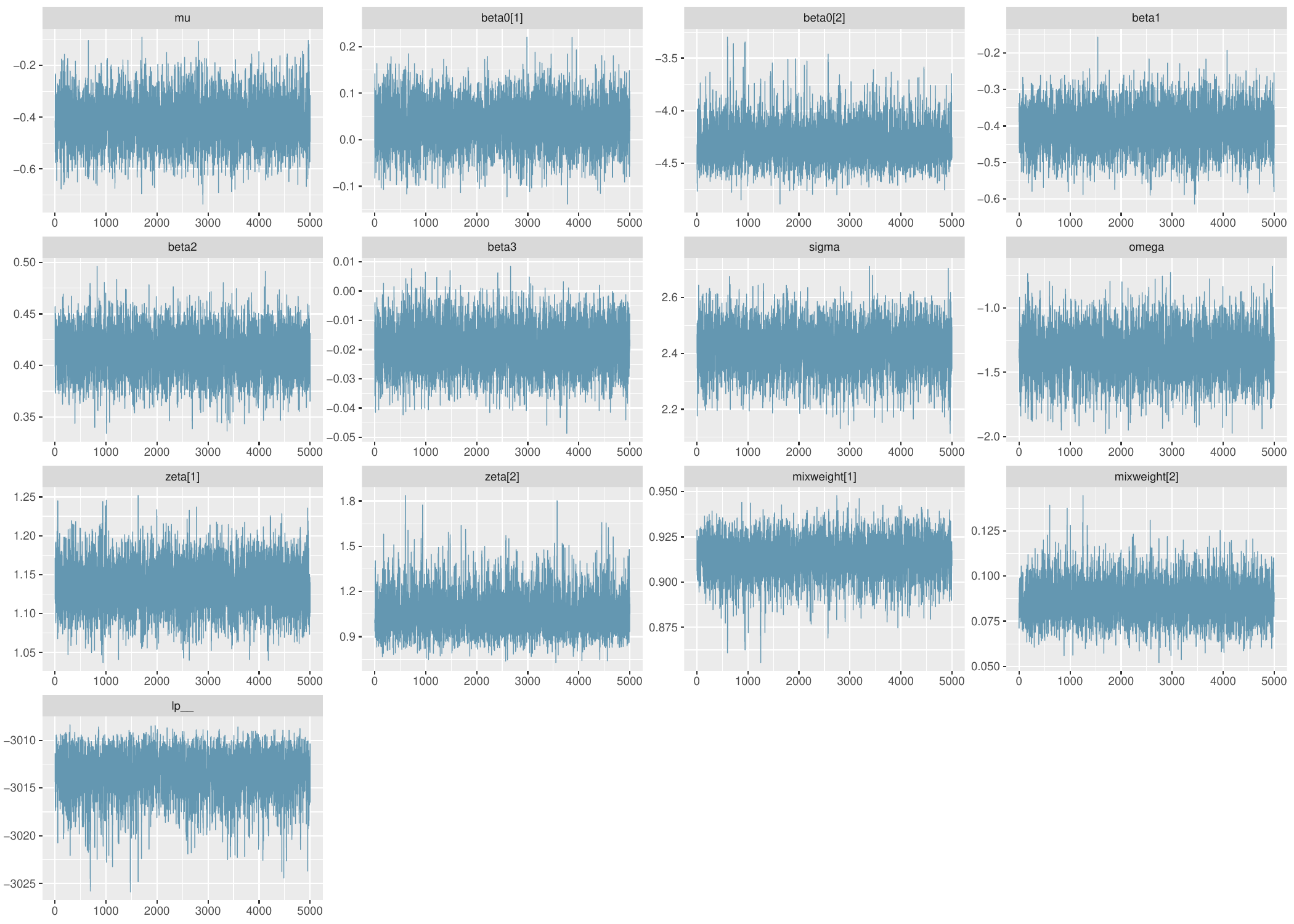}
    \caption{Trace plots from the complete case data BsNmN model analysis.}
    \label{fig:tracecomplete}
\end{figure}

\begin{figure}
    \centering
    \includegraphics[width=14cm]{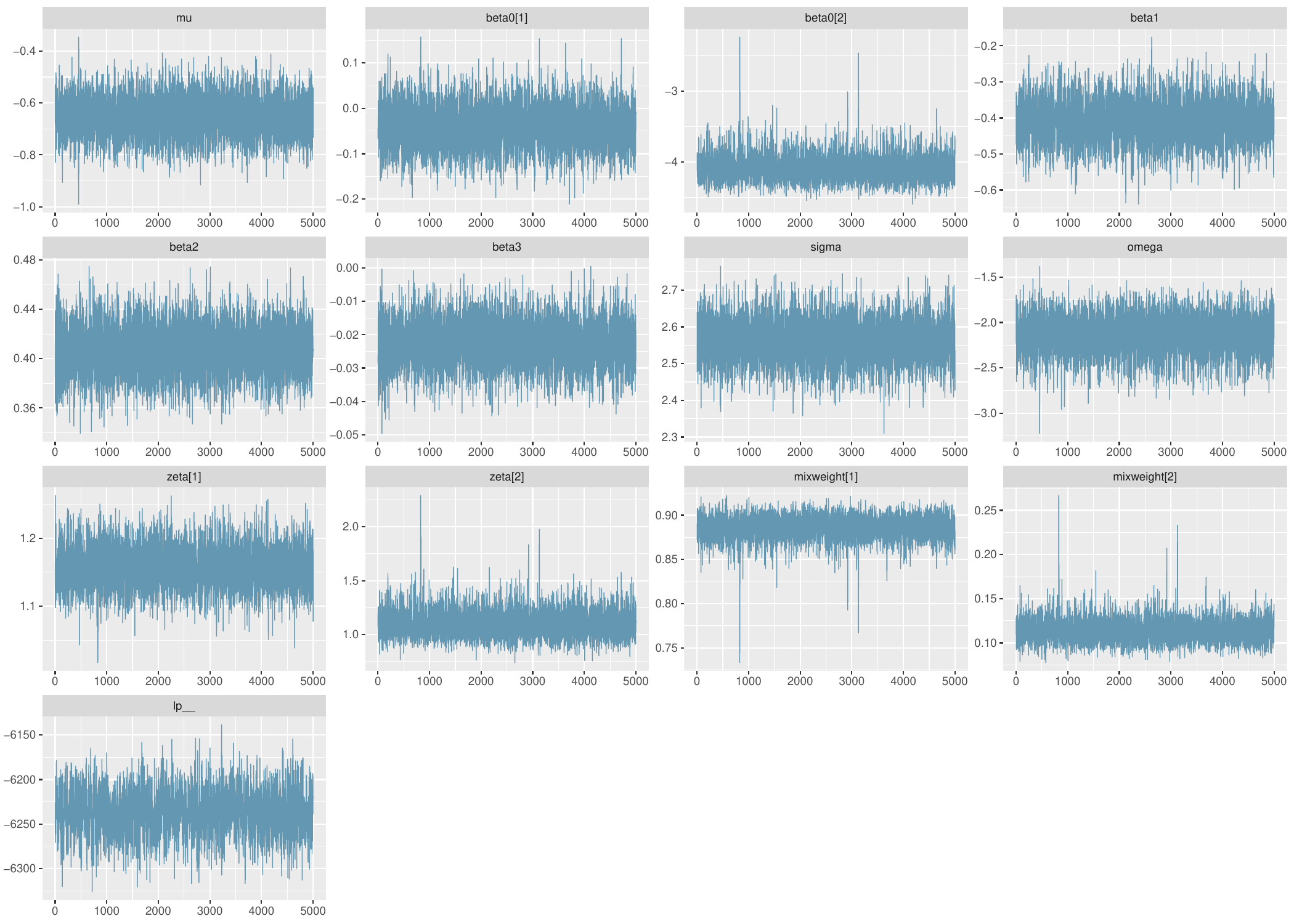}
    \caption{Trace plots from the entire dataset BsNmN model analysis.}
    \label{fig:tracefull}
\end{figure}

\end{footnotesize}

\end{document}